\documentclass[12pt]{article}
\usepackage{amsmath}
\usepackage{graphicx,psfrag,epsf}
\usepackage{enumerate}
\usepackage[colorlinks=true,linkcolor=blue,citecolor=blue,urlcolor=blue]{hyperref}
\usepackage{array}      
\usepackage{tabularx}    
\usepackage{adjustbox}  
\usepackage{rotating}
\usepackage{pdflscape}
\usepackage{booktabs}
\usepackage{pdflscape}
\usepackage{longtable}
\usepackage{xcolor}
\usepackage{natbib}
\usepackage{titletoc}

\newcommand{\blind}{0}
\DeclareRobustCommand{\likai}[1]{{\color{black}#1}}
\DeclareRobustCommand{\revtwo}[1]{{\color{black}#1}}

\begin{document}

\def\spacingset#1{\renewcommand{\baselinestretch}%
{#1}\small\normalsize} \spacingset{1}

%%%%%%%%%%%%%%%%%%%%%%%%%%%%%%%%%%%%%%%%%%%%%%%%%%%%%%%%%%%%%%%%%%%%%%%%%%%%%%

\if0\blind
{
  \title{\bf \likai{Persistence Paradox in Dynamic Science: Evidence from the Deep Learning Revolution}}
\author{Honglin Bao\ \thanks{These authors contributed equally to this work.
        Corresponding author: \textcolor{blue}{honglinbao@uchicago.edu}}
  \hspace{.2cm}\\
  Data Science Institute, University of Chicago\\
  Beichen Lu\footnotemark[\value{footnote}]\\
  School of Public Policy and Management, Tsinghua University\\
  Kai Li \\
  School of Information Sciences, University of Tennessee, Knoxville}
  \maketitle
} \fi

\if1\blind
{
  \bigskip
  \bigskip
  \bigskip
  \begin{center}
    {\LARGE\bf \likai{Persistence Paradox in Dynamic Science: Evidence from the Deep Learning Revolution}}
\end{center}
  \medskip
} \fi

\bigskip
\begin{abstract}
\likai{Persistence is often regarded as a virtue in science. In this paper, however, we challenge this conventional view by highlighting its contextual nature, particularly how persistence can become a liability during paradigm shifts. We focus on the deep learning revolution catalyzed by AlexNet in 2012. Analyzing the 20-year career trajectories of more than 5,000 scientists active in top machine learning venues during the preceding decade, we examine how their research focus and output evolved. We first uncover a dynamic period in which leading venues increasingly prioritized cutting-edge deep learning developments, displacing traditional statistical learning methods. Scientists responded to these changes in markedly different ways: those who were previously successful or affiliated with established teams adapted more slowly. Such persistence is positively associated with productivity but, after 2012, negatively associated with scientific impact. 
Most researchers, and the largest share of the field's output, cluster in a band of moderate persistence, pointing to a trade-off between output and impact, as well as to institutional frictions that make larger departures costly.
These conclusions are robust to alternative identification strategies and to competing explanations such as topic popularity premiums and survivorship bias. Taken together, our macro- and micro-level findings suggest that, in this case, a paradigm shift creates an opportunity structure by devaluing the very expertise that conferred incumbents' advantage in the first place.}
\end{abstract}
%that the rise of deep learning coincided with declining productivity among certain researchers, who were gradually supplanted by those forming new collaborations. This shift was driven by 
\noindent%
\vfill

\newpage
\spacingset{1} % DON'T change the spacing!
\section{Introduction}
\label{sec:intro}

Scientists are trained in specialized fields. While the ideal of the “Renaissance man” has long been celebrated in intellectual communities \citep{field2016unhelpful}, the scientific system is increasingly marked by specialization, leading to a growing “burden of knowledge,” which makes it challenging for any single individual to fully grasp all aspects of their domain \citep{jones2009burden, uzzi2013atypical, wu2022metrics}. This change could drive the increasing reliance on collaboration in the scientific system \citep{beaver1978studies, sonnenwald2007scientific} as well as the decreasing disruption in contemporary scientific research \citep{li2024displacing}. Most scientists tend to focus on a limited number of research topics, despite recent trends toward exploration \citep{zeng2019increasing}.

Against this backdrop, the debate between persistence and flexibility becomes central to understanding a researcher’s career trajectory. On the one hand, flexibility has long been considered useful for successful scientific careers. Science is often described as an “endless frontier” \citep{bush1945science} requiring researchers to adapt to new questions, topics, methods, and data \citep{harford2011adapt, omenn2006grand, teece1997dynamic}. Collaboration, in particular, has proven to be a powerful mechanism for researchers to acquire the expertise needed to pivot their research directions \citep{zeng2022impactful, venturini2024collaboration}.

On the other hand, a contrasting view posits that researchers who concentrate intensively on a single niche and persistently build on their prior work are more likely to achieve long-term impact. Social systems, including scientific communities, are inherently resistant to change to some degree \citep{brandt2012social, kuhn1997structure}. Empirical evidence suggests that adaptive researchers or those without a clear disciplinary identity often face penalties in both the production and perception of their work \citep{hill2025pivot, leahey2017prominent, schilling2011recombinant, wagner2011approaches}. Furthermore, although these researchers usually achieve higher long-term impact, their work may receive delayed recognition \citep{zhang2024delayed, leahey2017prominent}. Recent works have uncovered the career-dependent nature of persistence, noting that frequent topic switching negatively affects early-career researchers’ productivity but positively impacts productivity later in their careers \citep{zeng2019increasing}. Slightly shifting one’s research focus can be beneficial for postdocs entering the faculty job market, even if this strategy is not universally adopted by all researchers \citep{duan2025postdoc}.

\likai{In this paper, we contribute to this debate by examining how persistence relates to productivity and impact during \textit{a paradigm shift} in a field whose frontier is well represented in its leading publication venues. Our analysis focuses on a transformative period in which deep neural networks rose and displaced relatively traditional statistical machine learning methods\footnote{We discuss this history in the next section.}, following the publication of AlexNet \citep{alom2018history, krizhevsky2012imagenet}. We scope our claims accordingly: the setting is a high-velocity, conference-centric paradigm shift, and our evidence speaks to how the persistence--adaptation trade-off operated under these conditions rather than to scientific careers in general. Within this setting, persistence is positively associated with productivity but, after 2013, negatively associated with citation impact, a pattern we term the \textit{rigidity penalty}. We also find that most scientists clustered in a band of moderate overlap with their prior work, even though per-author impact kept rising as overlap fell further. \revtwo{Because greater overlap is also associated with higher productivity, this clustering may reflect a trade-off between output and impact, as well as the institutional frictions that make larger departures costly.}}

\likai{Our contribution builds on March's \citeyearpar{march1991exploration} exploration--exploitation framework, which describes a fundamental tension between refining existing competencies and venturing into uncertain new territory. This framework, and the literature it has inspired, largely treats the trade-off as a gradual calculus: the costs of exploitation accumulate slowly as a research agenda matures, so persistence is locally rational most of the time \citep{march1991exploration, jones2009burden, foster2015tradition}. Our findings suggest that a fast paradigm shift such as the deep learning revolution creates a qualitatively different condition. The value of accumulated expertise depreciates rapidly, the field's reward structure is reorganized around the emerging paradigm, and the costs of exploitation rise far more steeply than in periods of normal science, so that persistence can turn from an asset into a liability \revtwo{for impact, even as it continues to sustain output}. This depreciation also reshapes who has opportunities in the field. \citet{azoulay2019does} show that outsiders move into a field after its star scientists exit through death. Our results suggest that a comparable opening for lower-status scientists can arise without anyone exiting: when incumbents remain active but do not adapt, the shift itself erodes the expertise on which their advantage rested.}

The remainder of this paper is organized as follows. In Section~\ref{section2}, we provide a concise historical overview of deep learning, emphasizing the period that forms the focal point of our study. Section~\ref{section3} reviews related work and theoretical perspectives that underpin and motivate our analysis. In Section~\ref{section4}, we describe the datasets employed in our research. Section~\ref{section6} investigates the adaptation mechanisms and their relationship to researchers’ success. We conclude with final remarks in Section~\ref{section7}.

\section{A Brief History of Deep Learning}
\label{section2}

\likai{Over the past three decades, machine learning has shifted from relatively traditional statistical methods toward deep learning models: multi-layered architectures with very large numbers of parameters whose scalability and performance have come to dominate the field. Although the transition drew on many research contributions and institutional changes, it has a widely acknowledged starting point: AlexNet \citep{krizhevsky2012imagenet}, presented at the Neural Information Processing Systems conference in December 2012, which achieved a breakthrough error rate on ImageNet and which Geoffrey Hinton, Yoshua Bengio, and Yann LeCun later identified in their \textit{Nature} review as the catalyst of the deep learning revolution \citep{lecun2015deep}. We therefore treat 2013 as the first year of the paradigm shift. The shift was largely unanticipated by the mainstream machine learning community: deep neural networks had been discussed for decades but were widely regarded as computationally impractical (a problem later addressed by GPU acceleration) and as lacking theoretical elegance \citep{alom2018history}.}

\likai{Adoption of the new paradigm was not uniform across the community. In 2013, LeCun, Bengio, and others founded the International Conference on Learning Representations (ICLR), a venue dedicated to deep learning. Within a few years, ICLR became one of the three most impactful machine learning conferences, alongside the International Conference on Machine Learning (ICML) and Neural Information Processing Systems (NeurIPS, known as NIPS before 2018), both established in the 1980s\footnote{Accessed on Aug 27, 2024, from \href{https://scholar.google.es/citations?view_op=top_venues&hl=en&vq=eng_artificialintelligence}{Google Scholar's impact rank of AI venues}.}. The three venues mutually acknowledge each other's prominence: ICLR, for example, requires potential reviewers to have at least one accepted paper at a prior ICLR, NeurIPS, or ICML\footnote{https://iclr.cc/Conferences/2025/CallForPapers. Accessed on Dec 31, 2024. In computer science, papers in a conference proceeding are considered formal publications.}. We treat these venues as layers of proximity to the new paradigm. ICLR is the venue closest to deep learning by design. ICML and NeurIPS form a second layer: as leading venues predating the revolution, they published much of the work on which it built, and many of ICLR's founders were central contributors to them. Other forums in artificial intelligence, statistics, natural language processing, and computer vision also publish high-quality deep learning research but are more distant from its core. Appendix~\ref{appendix:distance} provides empirical evidence for these distances.}

\section{\texorpdfstring{Institutional and Individual Responses \\ to New Paradigms}{Institutional and Individual Responses to New Paradigms}}
\label{section3}
\subsection{Institutional Responses to New Paradigms}

The scientific enterprise oscillates between periods of stability and episodes of disruptive change. During periods of “normal science,” research communities adhere to shared paradigms and incrementally extend them, creating a stable hierarchy of knowledge and authority \citep{kuhn1997structure}. Over time, however, unresolved anomalies accumulate and radical new ideas – often initially developed at the margins – precipitate a crisis leading to a paradigm shift \citep{azoulay2014matthew, hull1978planck, bao2025division}. Such disruptive paradigm shifts can upend existing hierarchies and reward structures in science. Established leaders and institutions, heavily invested in the old paradigm, may resist the new one even as evidence mounts. In effect, the hierarchy is overturned as younger scholars embracing the new framework eventually replace the old guard. \likai{In addition, more recent evidence has shown that when a dominant figure in a field is removed or a prevailing theory collapses, an opening for innovation by younger generations can emerge \citep{azoulay2014matthew, azoulay2019does}, potentially allowing diverse approaches to flourish until the field’s overall productivity recovers and even exceeds its previous level.} This supports a variant of Planck’s principle \citep{hull1978planck}: rather than age alone, it is often the position and influence of incumbents that shape inertia. As long as a dominant elite sets research agendas (consciously or not), outsiders struggle to introduce disruptive innovations. When that hold is broken – whether by generational turnover, a scientific crisis, or external shock – hierarchical barriers give way to new ideas and players \citep{bergek2013technological, konig2012inertia, siler2024gerontocracy, cattani2017deconstructing}. 

Institutional inertia - including conferences, journals, universities, and funding bodies – plays a critical role in mediating the tension between persistence and adaptation during paradigm shifts. These institutions often embody the prevailing consensus of the current paradigm and therefore tend to respond conservatively to radically new research \citep{lane2022conservatism}, while being more receptive to innovations that align closely with existing agenda \citep{yan2024cultural, yan2025measuring}. For instance, elite journals and conference program committees are typically staffed by established experts whose standards were forged under the prevailing paradigm. Through the norm of organized skepticism, they subject unconventional findings to especially stringent scrutiny \citep{merton1968matthew, lane2022conservatism, teplitskiy2022novel}. This rigorous peer review serves as quality control, ensuring that only well-substantiated paradigm-challenging results enter the literature. However, \likai{it also makes such venues structurally less equipped to evaluate work that crosses paradigm boundaries, because the growing burden of knowledge \citep{jones2009burden} narrows the expertise any single reviewer can bring to bear}, as reviewers demand extensive evidence to be convinced of claims that contradict textbook knowledge \citep{teplitskiy2022novel, azoulay2011incentives}. Reviewers even care more about how the work is positioned and connected with prior literature than its substantive contribution \citep{bao2025language, strang2015revising}. \likai{More generally, conformity to established research strategies tends to be rewarded with reliable, if modest, recognition, whereas departures from them carry higher risk \citep{foster2015tradition, uzzi2013atypical}.} In the early phase of a paradigm shift, researchers aiming to publish paradigm-breaking studies may seek out alternative outlets. We often observe the emergence of new journals, special issues, or conference tracks dedicated to the innovative approach, providing a collegial arena for like-minded scholars to develop the paradigm. Such specialized forums reduce the penalization of deviant ideas and allow the new paradigm to build credibility through the accumulation of results. Over time, if the paradigm gains adherents and empirical support, mainstream institutions adapt \citep{bao2025division}. \likai{For example, prestigious journals start accepting (and even soliciting) papers from the new paradigm \citep{teplitskiy2022novel, azoulay2025does, azoulay2011incentives, frankenhuis2023strategic}, and major conferences incorporate the once-controversial topic into their core dialogue.}

\likai{A parallel dynamic operates in research funding, though its institutional logic differs from that of publication venues. Journals and conferences can signal receptiveness to a new paradigm relatively quickly through editorial appointments and acceptance patterns; funding agencies adapt more slowly, as their review panels, portfolio commitments, and grant cycles are organized around longer time horizons and more entrenched evaluation criteria \citep{araj2024elements, azoulay2025does}. Empirical studies of grant evaluation find that review panels systematically discount the novelty of proposed research, favoring projects that extend established lines of inquiry over those that depart from them \citep{boudreau2012novelty, ayoubi2021does, shin2022scientific} — a conservatism that reflects not individual bias so much as the structural difficulty of evaluating work that lacks an established community of qualified reviewers. The result is that funding institutions tend to lag behind publication venues in recognizing and rewarding emergent paradigms, creating an uneven institutional landscape during periods of transition. Publication venues, by contrast, can reward novelty: journal peer review has been found to favor novel manuscripts, and editors select on novelty even conditional on reviewer scores \citep{teplitskiy2022novel}. A novelty premium at the publication stage can thus coexist with a novelty penalty in funding.}

\likai{These institutional dynamics bear directly on the individual trade-off discussed next. That scientific institutions are conservative is well established; less appreciated is that different institutions are conservative at different rates. Because publication venues can move toward a new paradigm faster than funding agencies, a scientist during a paradigm shift may be rewarded for adaptation by one evaluation system and penalized for it by another. This likely also happens across venues at different distances from the paradigm shift, with leading venues acting as the most sensitive indicators of new evaluation practices and the primary drivers of them. Relatedly, whereas \citet{azoulay2019does} show that the exit of dominant incumbents opens space for outsiders, a paradigm shift may generate a comparable opportunity structure without anyone exiting, by devaluing the expertise that conferred incumbents' advantage in the first place. We develop both points in the next subsection.}

\subsection{Individual Responses to New Paradigms} \likai{Individual scientists face what \citet{march1991exploration} termed the fundamental tension between \textit{exploitation}, i.e., refining established expertise, and \textit{exploration}, i.e., venturing into uncertain territory for potential breakthroughs. This framework, and the large body of literature it has inspired, generally treats the trade-off as a continuous calculus in which the costs of exploitation accumulate gradually and persistence is often the locally optimal response. Scientists are indeed more likely to pursue incremental, ``normal science'' projects than to tackle novel and high-risk questions \citep{rzhetsky2015choosing, foster2015tradition}, a tendency reinforced by institutional reward structures that favor cumulative productivity and established reputations \citep{wahid2024empirically}, as reviewed in the previous section. Persistence is thus not only a product of individual risk aversion or cognitive rigidity; it is also a rational response \revtwo{to evaluation systems that reward the steady output persistence sustains, and} to a funding environment that systematically penalizes novelty \citep{boudreau2012novelty, ayoubi2021does, shin2022scientific}. More broadly, \citet{fortunato2018science} provide a framework for studying the dynamics of scientific careers and fields that underscores how both individual strategies and field-level forces jointly shape research trajectories — a perspective that informs our dual macro- and micro-level analysis.}

\likai{Paradigm shifts, however, represent a qualitatively different condition. Rather than eroding gradually, the value of accumulated expertise can depreciate abruptly and discontinuously as the field's reward structure reorganizes around an emerging framework. This is the puzzle our case addresses. Existing frameworks predict that persistence is locally rational under stable conditions \citep{march1991exploration, jones2009burden, foster2015tradition} and that the exit of incumbents opens a field to outsiders \citep{azoulay2019does}. They are largely silent, however, on how the returns to persistence change when the paradigm itself shifts while incumbents remain active. The deep learning revolution is particularly well suited to isolating the effect, because the adaptation of the evaluating venues themselves is directly observable (Section~\ref{6.1}).}

\likai{These cross-pressures have elicited diverse and often nuanced responses from researchers. Prior work suggests that scientists' new work tends to emerge not from a complete abandonment of prior expertise but from a balance between continuity and change: modest rather than wholesale departures from prior research strategies are the most common \citep{uzzi2013atypical, foster2015tradition}. These findings point to a more complex relationship between persistence and adaptation. Complete rigidity forecloses access to the new paradigm's rewards, whereas complete abandonment of prior expertise sacrifices credibility and transferable skills and, given the growing burden of knowledge \citep{jones2009burden}, incurs retraining costs that most scientists cannot absorb quickly \revtwo{and are paid first in lost output}. This literature therefore suggests that \textit{moderate adaptation}---selectively pivoting toward emerging trends while retaining connections to prior expertise---is a natural strategy. Whether it also maximizes individual returns, or merely describes where most scientists end up, is an empirical question that we test directly in Section~\ref{5.2.3}. \revtwo{Because persistence and adaptation may reward different outcomes, the answer may depend on which return is considered.}}

Cognitive and social factors also determine why some scientists adapt while others persist. Cognitive lock-in and expertise traps can make a veteran researcher deeply invested in a particular paradigm less likely to “see” merit in new approaches \citep{bao2024simulation}. By contrast, newcomers or those with interdisciplinary training might more readily apply fresh frameworks. Age is also a predictor of openness \citep{azoulay2019does}. In some historical cases – for example, the acceptance of plate tectonics in geology – senior scientists actually adopted the new paradigm faster than their juniors \citep{diamond1980age, messeri1988age}. Such cases suggest that established experts can and do pivot under the right conditions, perhaps due to broader experience or fewer career risks at late stages. Other evidence shows that across all fields, time periods, and impact levels, scientists increasingly rely on older ideas and references as their careers progress \citep{cui2022aging}. Nonetheless, the costs of switching fields or approaches (in time required to retrain, loss of status in one’s old community, uncertainty of recognition in the new area) mean that many researchers only make bold shifts when compelled by clear evidence or lack of progress in their current line. Those who do successfully adapt often leverage transferable skills to ease the transition – for instance, applying a known method to a new domain \citep{ao2025researching}. 

\likai{These dynamics suggest} that low-status scientists are more inclined to change their research directions during periods of scientific disruption. Limited access to resources, recognition, and opportunities within the existing paradigm creates structural constraints that may push low-status scientists to explore alternative paths \citep{bhatt2020neither, bhatt2022language, goldberg2016fitting, srivastava2018enculturation}. Disruptive events, which destabilize established hierarchies, may offer unique opportunities for these scientists to reposition themselves. By breaking from dominant paradigms, they can seek new avenues for development, effectively leveraging disruption to challenge the entrenched status quo \citep{zhang2021shaking}, while successful scientists are less likely to adapt because they are reluctant to let go of their existing achievements. \likai{This mechanism is also echoed by \citet{azoulay2019does}, who show that the removal of a field's apex scientists through death opens space for new entrants and alternative approaches. In our setting, the relevant dynamic is not exit but \textit{persistent presence}: established scientists who remain active but do not adapt may produce a similar opening for lower-status peers, as their continued publication in familiar research directions yields diminishing citation returns in a field whose evaluation criteria have shifted.} Low-status scientists, by comparison, may experience a heightened motivation to adjust their research agendas during these periods, as they face fewer risks associated with deviating from the dominant scientific order and they may gain status during the process \citep{marantz2024changing}.

Conversely, \likai{an alternative} perspective posits that high-status scientists, by virtue of their abundant resources and influence, are more likely to first adapt during periods of scientific paradigm shifts. These scientists are often positioned at the forefront of their fields, leveraging their significant access to research funding, cutting-edge experimental infrastructure, and extensive academic networks to adapt to new paradigms \citep{azoulay2014matthew, merton1968matthew}. ICLR illustrates how prominent researchers in machine learning mobilized their resources and networks to institutionalize and promote advancements in deep learning \citep{lecun2015deep}. In addition to the prior status of scientists, the collaborator network is also likely to have a significant influence on an author's research adaptation. Strong collaborators may lead to greater changes in their research.

Research is a deeply social process: new ideas spread through formal and informal networks (colleagues, co-authors, mentors, conference communities), and a researcher’s status and position in these networks can either constrain or enable adaptation. Overall, at the micro-level, a scientist’s decision to persist or pivot emerges from the interaction between collaborations, personal curiosity, risk tolerance, career incentives, and the perceived credibility of the emerging paradigm.

\section{Dataset}
\label{section4}
To examine how science adapts during a transformative era, we focus on researchers active before the deep learning boom in 2013. We begin by collecting all 22,651 papers published at NeurIPS and ICML in the ten years before and after 2013 (2003-2022) and 5,163 ICLR papers published during 2013-2022, capturing shifts in the machine learning landscape. We also assemble a cohort of 5,359 scholars who published at least one paper in ICML or NeurIPS between 2003 and 2012—i.e., the decade preceding the deep learning surge—and construct a panel dataset tracking their annual research output at those same conferences from 2003 through 2022 (ten years before and after the deep learning boom). 

\begin{figure}[htbp]
    \centering
    \includegraphics[width=0.5\textwidth]{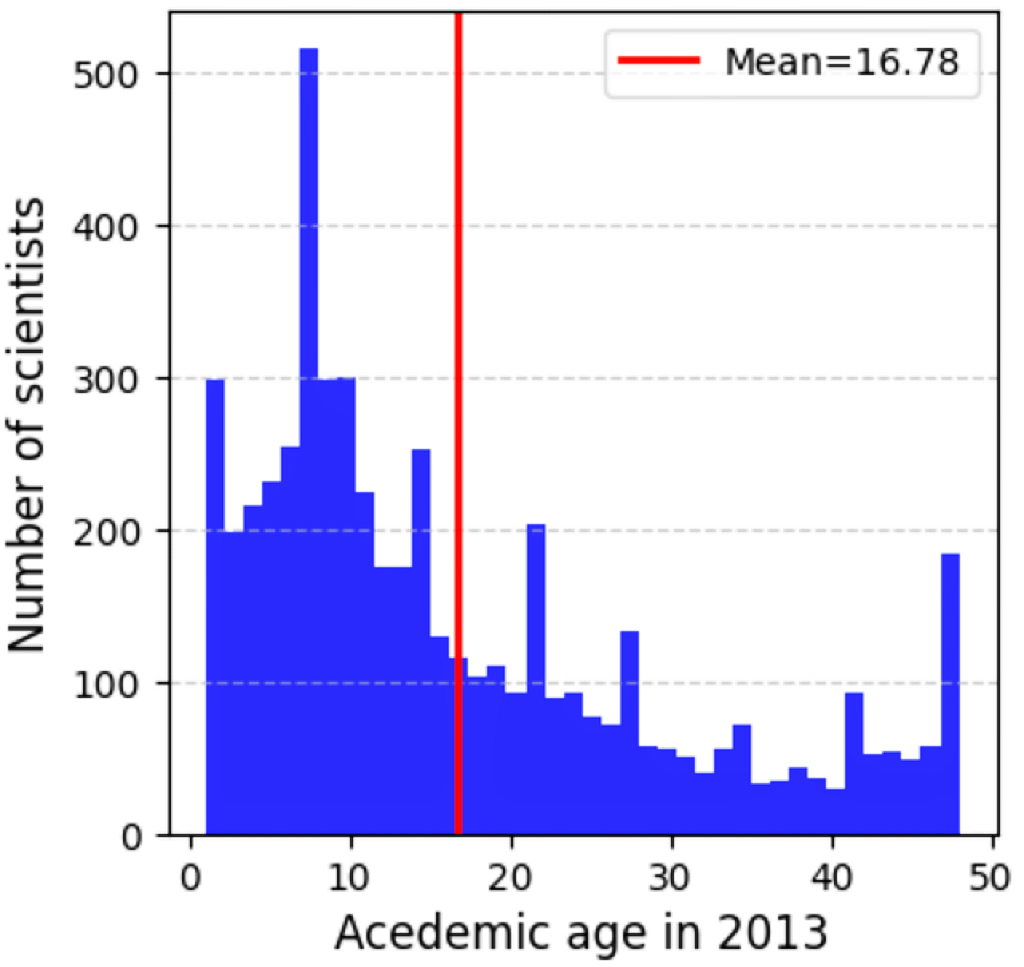} % full width
    \footnotesize
    \caption{The seniority of scientists at the time of the machine learning paradigm shift. \likai{Academic age is the number of years between the scientist's first publication in OpenAlex and 2013. The sample is the cohort of 5,359 scientists with at least one ICML or NeurIPS paper in 2003--2012.}}
    \label{fig:age11}
\end{figure}

We did not use these authors' papers back to the earliest stages of machine learning in the 1980s. Given the generational turnover of scientists and the evolutionary nature of scientific progress, their papers between 2003 and 2012 are more representative of the pre-deep learning scientific community and the immediate reaction of the community to a paradigm shift. These scientists were relatively senior when they encountered the deep learning boom in 2013. We recorded their academic age—calculated as the difference between 2013 and the year of their first publication in the OpenAlex database—to represent their experience at the time of this paradigm shift (Figure \ref{fig:age11}) \footnote{This age is the academic seniority of a scientist. Note that in the following sections, we also have the "team age" (i.e., collaboration familiarity), defined as the average number of previous collaborations on ICML/NeurIPS papers between the focal scientists and their coauthors; see Section \ref{6.2}.}. On average, they had been active in the field for over a decade, having published in the most prestigious venues during the formative years of 2003–2012.

Our central research setting is the group of machine learning researchers who published in top-tier conferences (ICML and NeurIPS) over the decade leading up to the advent of deep learning. When deep learning began its explosive growth in 2013, would these researchers - if they continued pursuing their established topics and methods - still be able to publish at the same premier venues, given the rapid evolution of the field, which is arguably well reflected in these conferences \footnote{We have more discussions in Appendix \ref{appendix:distance}.}? Furthermore, how would their productivity and the impact of papers change compared to other papers appearing in ICML and NeurIPS during this transformative period?

The purpose of selecting this sample is two-fold: (a) ICML and NeurIPS are premier conferences with a plausibly high bar of quality and visibility. They are particularly relevant to the deep learning revolution and have the advantage of existing both before and after 2013, making them ideal as a reference for longitudinal analysis. (b) If a scientist lowers their standards for where they submit papers, their productivity will almost always increase, regardless of whether they actually adapt their work. In Appendix \ref{appendix:distance}, we analyze 579,973 papers from the Microsoft Academic Graph (MAG) dataset, categorized under "machine learning" or "deep learning," spanning the same 20-year period of deep learning formation and growth. Using \likai{SPECTER2 document embeddings}, we empirically demonstrate that, compared to ICML and NeurIPS, general machine learning publications are less closely tied to the rise of deep learning. In other words, these broader publications lack a consistent standard for “selecting” cutting-edge work, or their submitting authors are less sensitive to the change, limiting their ability to provide a clear analysis of the evolution of fields and scientists.

We would like to highlight two key aspects of research design. First, name disambiguation is particularly crucial in computer science papers due to the high prevalence of common Chinese names \citep{kim2023effect}. In our study, the scientists in our sample published at least one paper in ICML or NeurIPS between 2003 and 2012 - a period during which Chinese scientists were less represented in these conferences compared to their prevalence today. We use names, email addresses, and affiliations for initial name disambiguation. However, this approach is still limited, particularly in cases where individuals change their email addresses/institutions after graduating. To improve accuracy, we further utilize the Google Scholar API, Scholarly, for disambiguation. If two authors (with different email addresses or affiliations) have the same name, and their publications on Google Scholar indicate they are the same individual, we merge them into a single author record. \likai{A manual check of 100 randomly sampled authors yields a 96\% match rate. Appendix~\ref{subsec:name_disambiguation} describes the procedure and validation in full.}

Second, a significant confounder in our research design is the establishment of ICLR in 2013, which quickly rose to become a top-tier conference alongside ICML and NeurIPS. This development likely influenced the publishing intentions and patterns of our focus group in ICML and NeurIPS. \likai{Because the earliest ICLR authors were likely disproportionately drawn from the most adaptive members of our cohort, excluding ICLR could bias our estimates. In Appendix~\ref{appendix:alternatives}, we show that ICLR accounted for a modest share of the cohort's output in its early years, and we re-estimate the main model on an ICLR-inclusive panel that adds the cohort's ICLR papers; the main conclusions are unchanged.}

\section{Adaptation and Success}
\label{section6}
This section presents our analysis of the relationship between adaptation and success within the evolving academic landscape. We begin with a macro-level perspective on how the field is changing (Section \ref{6.1}), followed by a micro-level analysis (Section \ref{6.2}), which explores whether adaptation \likai{is associated with} success\revtwo{, measured separately as productivity and impact} (Section \ref{6.2.1})\revtwo{,} who is more likely to adapt (Section \ref{6.2.2})\revtwo{, and where scientists cluster relative to these two returns (Section~\ref{5.2.3})}. \likai{We then examine alternative explanations (such as the topic premium and survivorship bias) and complementary designs, including an event study, propensity score matching, and a quasi-experimental pre-shock trend analysis (Section~\ref{sec:alt}).} Finally, we summarize the key insights from our results (Section~\ref{6.3}).

\subsection{Macro-level Analysis}
\label{6.1}

How do ICML and NeurIPS capture the cutting edge of deep learning? ICLR was founded by the pioneers of deep learning to establish a dedicated venue for this rapidly growing field, making its research highly representative of deep learning. Using the TF-IDF (Term Frequency-Inverse Document Frequency) algorithm, we extracted the top ten keywords from each paper's abstract. TF-IDF is a statistical method widely used to evaluate how important a word is to a document (abstract) within a corpus of abstracts. The TF component measures how often a term appears in an abstract, capturing its relevance within that context. The IDF component, on the other hand, measures how unique or rare the term is across all abstracts in a year. By combining these two factors, TF-IDF assigns higher weights to terms that are frequent within one abstract but relatively rare across the entire corpus, allowing it to effectively highlight keywords that best represent the focal abstract's unique topics. This approach provided a robust representation of the topics dominating ICLR each year and allowed for meaningful comparisons with topics emphasized at ICML and NeurIPS. An example of an abstract and its extracted tokens can be found in Appendix \ref{word_table}. \likai{TF-IDF is used only for this venue-level topic comparison; the scientist-level persistence measure in Section~\ref{6.2} is based on SPECTER2 document embeddings.}

\begin{figure}[htbp]
    \centering
    \includegraphics[width=0.9\textwidth]{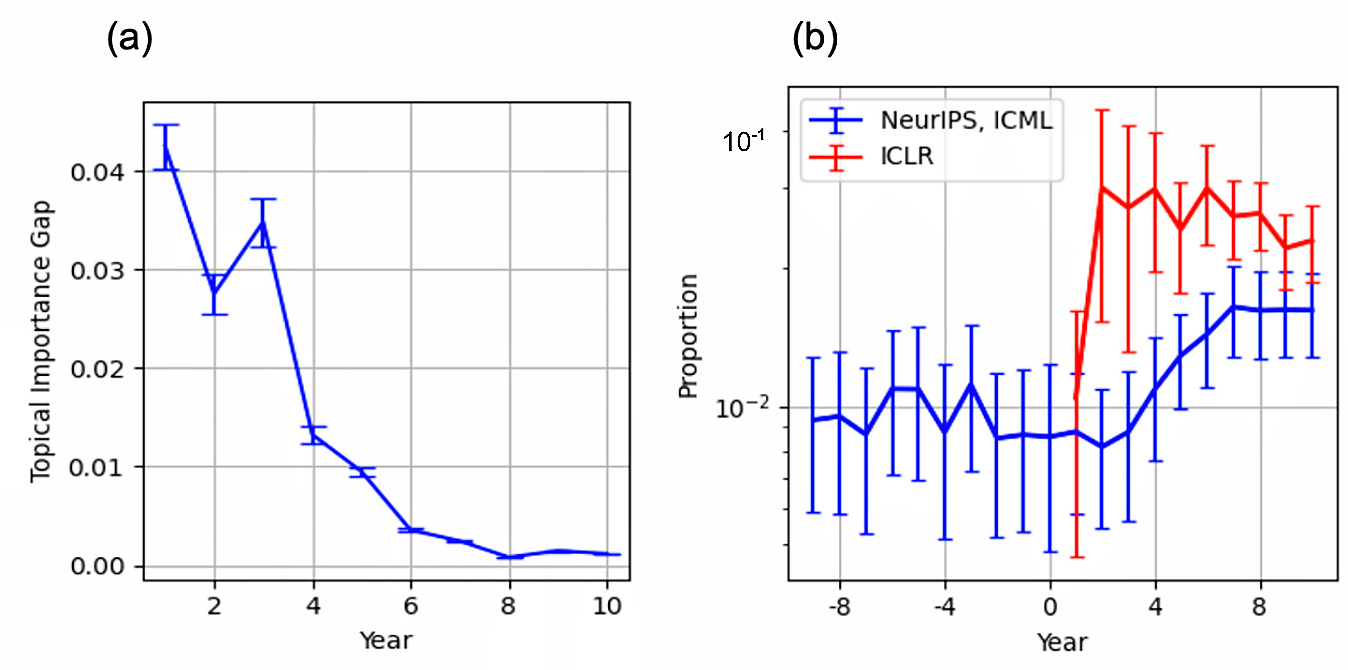} % full width
    \footnotesize
    \caption{Topical convergence between ICML, NeurIPS, and ICLR. Panel (a): the topical importance gap has been decreasing over the years. Panel (b): the ICLR-specific topics have been becoming more and more important in ICML and NeurIPS. \likai{Topics are the top-ten TF-IDF keywords extracted from each abstract. The topical importance gap is the absolute difference between a topic's frequency in ICML/NeurIPS abstracts and its frequency in ICLR abstracts in the same year, computed for every ICLR topic. ICLR-specific topics are the 50 topics with the largest such gap during 2013--2022.} Error bars represent 95\% confidence intervals.}
    \label{fig:age3}
\end{figure}

For every topic identified in ICLR in a given year, we calculated its frequency in ICML and NeurIPS abstracts from the same year (assigning a frequency of 0 if absent). We then computed the distribution of absolute differences in topic importance (frequency) between the ICML and NeurIPS corpus versus the ICLR corpus for that year. As shown in panel (a) of Figure \ref{fig:age3}, a clear trend emerges: the topics of the three conferences have become increasingly similar over time. Next, we identified the deep learning topics most underrepresented in ICML and NeurIPS, focusing on the top 50 topics with the largest differences in frequency between ICLR and ICML/NeurIPS during the first decade of the deep learning wave. These topics represent areas significantly more emphasized in ICLR than in the other two conferences. How have these topics evolved within ICML and NeurIPS? As shown in panel (b) of Figure \ref{fig:age3}, their importance began to rise following the onset of the deep learning wave. Moreover, the relative importance of these topics has converged among ICLR, NeurIPS, and ICML. These findings suggest that ICML and NeurIPS, either actively (through peer review and selection) or passively (through cross-conference submissions), have adapted to increasingly capture the cutting edge of deep learning.

\subsection{Micro-level Analysis}
\label{6.2}
\subsubsection{\texorpdfstring{\revtwo{Is adaptation associated with success, and of which kind?}}{Is adaptation associated with success, and of which kind?}}
\label{6.2.1}
In this section we study how adaptation \likai{is associated with} success at the scientist level.\\

\textbf{Dependent Variables (Academic Success)}:

\begin{itemize}
    \item \textbf{Productivity}: The number of papers published by a scientist in ICML and NeurIPS in a given year.
\item \textbf{Impact} (0-100): For each paper, we calculate its citation percentile rank among all papers published in the same conference that year. A scientist's annual impact is the average percentile rank of all their papers for that year.
\end{itemize}

\textbf{Independent Variables}:

\begin{itemize}
    \item \textbf{Research Persistence} (0-100): To quantify a scientist’s research persistence from year $T$ to $T$ + 1, we first gather all ICML/NeurIPS papers authored by the scientist up to and including year $T$. Next, we collect their ICML/NeurIPS papers from year $T$ + 1. To assess persistence, we compute the bootstrapped text similarity between these two corpora \likai{using titles and abstracts}. Since the sizes of the two corpora (before $T$ and $T$ + 1) may differ significantly—introducing potential bias (text similarity is typically lower when comparing a short text to a longer one)—we address this by matching their sizes. Specifically, we retain the smaller corpus and randomly sample a subset of texts from the larger corpus to match the size of the smaller corpus. \likai{In most author-years the focal-year corpus is the smaller one, so the procedure subsamples the prior corpus; this asymmetry is exactly what the size matching is designed to neutralize. We then calculate the embedding-based text similarity between the sampled subset and the smaller corpus using SPECTER2 \citep{singh-etal-2023-scirepeval}, a state-of-the-art scientific document encoder pre-trained on large-scale literature with citation-based contrastive learning, which to our knowledge is the most accurate embedding model currently available for representing scientific documents. The measure is therefore meaning-aware rather than lexical: a scientist who applies deep learning methods to a problem they previously studied with other methods will register as having moved away from their prior work.} This process is repeated 1,000 times, and the average similarity across iterations is used as the similarity score from year $T$ to $T$ + 1. A higher bootstrapped similarity score indicates greater persistence and less adaptation. Interestingly, there is generally low persistence (41 $\pm$ 1 on a 0-100 scale) observed among authors in the field of computer science. \likai{Because different text-representation algorithms can yield substantially different similarity scores for the same documents \citep{tshitoyan2019unsupervised}, we validate the measure against two alternatives in Appendix~\ref{persistence_m}: a five-year-window version that compares the focal year only with the scientist's papers from the preceding five years (correlation with the main measure $r=0.87$), and a topic-level measure based on a 30-topic LDA model that requires no size matching at all. Both reproduce the main results.}
    
\end{itemize}

\textbf{Control Variables}: 

We control for the prior performance of scientists and the contributions of their collaborators. The core idea of our analysis is that a scientist's success in a given year can be decomposed into three key components: (1) their individual ability to produce papers (measured by their prior performance), (2) the contributions of their collaborators (how many collaborators, how well the focal author knows them, and how good the collaborators are), and (3) the premium, or penalty, associated with research adaptation as determined by the publishing market, if any, which is our key independent variable. Adaptation could occur in part through collaboration. By controlling for components from collaboration networks, we isolate the additional \likai{association between adaptation and success, net of collaboration}.

\begin{itemize}
    \item \textbf{Previous Productivity/Previous Impact} (0-100): We use the author’s performance in the previous year to measure their prior foundation for publishing. 
\end{itemize}

The input from collaborators can be characterized by the number of collaborators, the strength of the focal author’s familiarity with them, the quality of those collaborators, and their potential interactions.

\begin{itemize}
    \item \textbf{Collaborator Count}: The total number of collaborators a scientist worked with on ICML/NeurIPS papers in a given year.
    \item \textbf{Collaboration Familiarity}: For each scientist, we track their unique collaborators in a given year and calculate the number of prior ICML/NeurIPS papers they coauthored with these collaborators (dating back to the inception of these conferences: ICML 1980 and NeurIPS 1987). The scientist’s collaboration familiarity with their coauthors for that year (i.e., team age) is defined as the average number of prior collaborations with their current collaborators.
    \item \textbf{Collaborator Status} (0-100): Collaborator status is measured as the average status of all collaborators in year $T$. The status of a collaborator in year $T$ is determined by their percentile rank on the cumulative citations of produced ICML and NeurIPS papers before year $T$ among all authors.
\end{itemize}

\begin{figure}[htbp]
    \centering
    \includegraphics[width=0.95\textwidth]{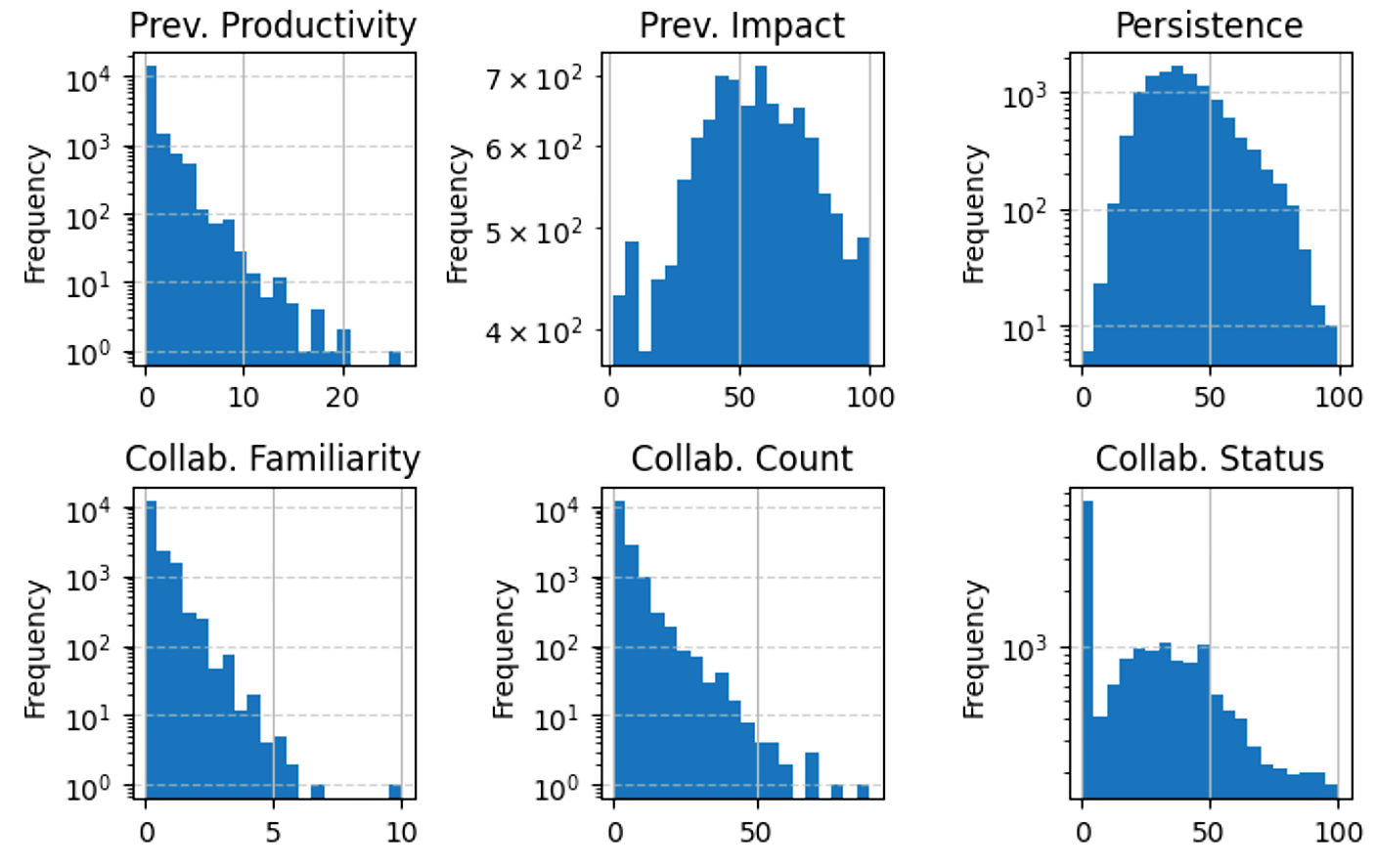} % full width
    \footnotesize
    \caption{Descriptive statistics for variables used in the regression. \likai{Productivity: number of papers in a year. Impact: mean within-conference-year citation percentile of those papers (0--100). Persistence: bootstrapped SPECTER2 similarity between a scientist's papers in a year and their prior papers (0--100). Collaborator count: number of coauthors in a year. Collaboration familiarity (team age): mean number of prior papers coauthored with current coauthors. Collaborator status: mean percentile rank of coauthors' cumulative citations. Axes for long-tailed variables are on a log scale.}}
    \label{fig:des}
\end{figure}

The relationship between variables of persistence, prior status, academic age, and collaboration network features is shown in Appendix \ref{appendix:rela}. Descriptive statistics for the variables presented can be found in Figure \ref{fig:des}. It reveals a long-tail distribution in productivity and collaboration networks.

\vspace{5mm}
\textbf{Regression}:

We track the cohort of scientists over 20 years using a regression with two-way fixed effects:

\begin{align*}
Impact/Productivity_{i,t} =\\
& \beta_1 \times \mathit{Research\ Persistence}_{i,t} 
+ \beta_2 \times \mathit{Research\ Persistence}_{i,t} \times \mathit{Post\ 2013}_t \\
& + \beta_3 \times \mathit{Prior\ Impact/Prior\ Productivity}_{i,t} \\
& + \beta_4 \times \mathit{Collaborator\ Familiarity}_{i,t} \\
&+ \beta_5 \times \mathit{Collaborator\ Count}_{i,t} \\
&+ \beta_6 \times \mathit{Collaborator\ Status}_{i,t} \\
& + \mathit{Potential\ Interactions}(\mathit{Collaborator\ Familiarity}_{i,t},\ 
\mathit{Count}_{i,t},\ \mathit{Status}_{i,t}) \\
& + \alpha_i + \gamma_t + \varepsilon_{i,t}
\end{align*}

\noindent
Where $\beta_4$ -- $\beta_6$ terms represent the input from the collaborator network. 
$\varepsilon_{i,t}$ is the error term. 
Author fixed effects $\alpha_i$ are used to capture author-level characteristics, such as the scientist's age, motivation, and learning ability when they face paradigm shifts. 
Year fixed effects $\gamma_t$ capture temporal trends and external factors that may influence publication outcomes in a given year, such as shifts in funding priorities, technological advancements, or broader disciplinary trends. 
These fixed effects help isolate variations due to the research environment from those attributable to individual scientists.

The model includes research persistence ($\beta_1$) and its interaction with the post-deep learning era ($\beta_2$) to examine how a scientist's \likai{adaptation of their research is associated with} outcomes in the context of significant technological shifts. ${Research\ Persistence}_{i,t}$ captures the scientist $i$'s adaptation and persistence from the period before year $T$ to year $T$.
These terms enable us to assess whether adaptivity \likai{is associated with} additional benefits, or penalties, following the advent of deep learning. \likai{Persistence is chosen by scientists and may co-move with unobserved, time-varying factors, most importantly the declining popularity of the topics they work on. Section~\ref{sec:alt} and Appendices~\ref{survivorship}--\ref{causal} report analyses that address the main threats to interpretation, including alternative mechanisms (such as survivorship bias and topic premium) and complementary identification including an event study, propensity score matching, and a quasi-experimental pre-shock trend analysis.}

In our regression results shown in Tables \ref{impact_table} and \ref{prod_table} (models 4, 5, and 6) we also tried models including interactions between features of the coauthor network in year $T$, 
\textit{Familiarity $\times$ Count}, 
\textit{Familiarity $\times$ Status}, 
\textit{Status $\times$ Count}, and 
\textit{Status $\times$ Count $\times$ Familiarity}, 
to control the potential synergistic effects of the familiarity of coauthors, coauthor status, and coauthor count on scientific outcomes. 
In all regressions standard errors are clustered at the author level.

\begin{landscape}
\begin{center}
\footnotesize
\begin{longtable}{@{\extracolsep{5pt}}lcccccc}
\caption{Dependent variable: impact} \label{impact_table}\\
\hline \hline
& \multicolumn{6}{c}{\textit{Dependent variable: impact}} \\
\cline{2-7}
& Model 1 & Model 2 & Model 3 & Model 4 & Model 5 & Model 6 \\
& (persistence) & (author) & (collab. network) & & & \\
\hline
\endfirsthead

\multicolumn{7}{c}%
{{\bfseries \tablename\ \thetable{} -- continued from previous page}} \\
\hline
& Model 1 & Model 2 & Model 3 & Model 4 & Model 5 & Model 6 \\
& (persistence) & (author) & (collab. network) & & & \\
\hline
\endhead

\hline \multicolumn{7}{r}{{Continued on next page}} \\
\endfoot

\hline
\endlastfoot

post 2013$\times$persistence & -0.163$^{***}$ & -- & -- & -0.204$^{***}$ & -0.177$^{***}$ & -0.177$^{***}$ \\
& (0.049) & -- & -- & (0.051) & (0.052) & (0.052) \\
persistence & 0.036 & -- & -- & 0.004 & -0.002 & -0.002 \\
& (0.045) & -- & -- & (0.046) & (0.046) & (0.046) \\
prev. impact & -- & -0.070$^{***}$ & -- & -0.080$^{***}$ & -0.080$^{***}$ & -0.080$^{***}$ \\
& -- & (0.014) & -- & (0.013) & (0.013) & (0.013) \\
collab. familiarity & -- & -- & -4.172$^{***}$ & -3.483$^{***}$ & -2.028$^{*}$ & -2.096 \\
& -- & -- & (0.590) & (0.638) & (1.034) & (1.148) \\
collab. status & -- & -- & 0.116$^{***}$ & 0.116$^{***}$ & 0.129$^{***}$ & 0.128$^{***}$ \\
& -- & -- & (0.013) & (0.015) & (0.021) & (0.022) \\
collab. count & -- & -- & 0.297$^{***}$ & 0.450$^{***}$ & 0.711$^{***}$ & 0.702$^{***}$ \\
& -- & -- & (0.063) & (0.071) & (0.111) & (0.149) \\
familiarity$\times$status & -- & -- & -- & -- & -0.006 & -0.004 \\
& -- & -- & -- & -- & (0.017) & (0.022) \\
familiarity$\times$count & -- & -- & -- & -- & -0.393$^{**}$ & -0.366 \\
& -- & -- & -- & -- & (0.121) & (0.270) \\
status$\times$count  & -- & -- & -- & -- & -0.003 & -0.003 \\
& -- & -- & -- & -- & (0.003) & (0.004) \\
familiarity$\times$count$\times$status & -- & -- & -- & -- & -- & -0.001 \\
& -- & -- & -- & -- & -- & (0.006) \\
\hline
two-way fixed effects & Y & Y & Y & Y & Y & Y \\
Observations & 11,347 & 11,159 & 16,294 & 11,016 & 11,016 & 11,016 \\
N. of groups & 2,794 & 2,754 & 5,227 & 2,739 & 2,739 & 2,739 \\
$R^2$ & 0.004 & 0.005 & 0.017 & 0.029 & 0.031 & 0.031 \\
Residual Std. Error & 1.433 & 1.664 & 3.154 & 3.965 & 4.093 & 4.094 \\
F Statistic & 16.323$^{***}$ & 43.563$^{***}$ & 64.564$^{***}$ & 41.696$^{***}$ & 29.669$^{***}$ & 26.701$^{***}$ \\
\hline
\multicolumn{7}{r}{\textit{Note:} $^{*}$p$<$0.05; $^{**}$p$<$0.01; $^{***}$p$<$0.001} \\
\end{longtable}
\end{center}
\end{landscape}

\begin{landscape}
\begin{center}
\footnotesize
\begin{longtable}{@{\extracolsep{5pt}}lcccccc}
\caption{Dependent variable: productivity} \label{prod_table}\\
\hline \hline
& \multicolumn{6}{c}{\textit{Dependent variable: productivity}} \\
\cline{2-7}
& Model 1 & Model 2 & Model 3 & Model 4 & Model 5 & Model 6 \\
& (persistence) & (author) & (collab. network) & &  & \\
\hline
\endfirsthead

\multicolumn{7}{c}%
{{\bfseries \tablename\ \thetable{} -- continued from previous page}} \\
\hline
& Model 1 & Model 2 & Model 3 & Model 4 & Model 5 & Model 6 \\
& (persistence) & (author) & (collab. network) & &  & \\
\hline
\endhead

\hline \multicolumn{7}{r}{{Continued on next page}} \\
\endfoot

\hline
\endlastfoot

post 2013$\times$persistence  & 0.037$^{***}$ & -- & -- & 0.009$^{***}$ & 0.005$^{**}$ & 0.005$^{**}$ \\
& (0.004) & -- & -- & (0.002) & (0.002) & (0.002) \\
persistence & 0.031$^{***}$ & -- & -- & 0.021$^{***}$ & 0.023$^{***}$ & 0.023$^{***}$ \\
& (0.002) & -- & -- & (0.002) & (0.002) & (0.002) \\
prev. productivity & -- & 0.376$^{***}$ & -- & 0.034 & -0.017 & -0.010 \\
& -- & (0.034) & -- & (0.021) & (0.017) & (0.017) \\
collab. familiarity  & -- & -- & 0.110$^{***}$ & -0.028 & -0.270$^{***}$ & -0.528$^{***}$ \\
& -- & -- & (0.019) & (0.021) & (0.050) & (0.073) \\
collab. status & -- & -- & -0.001$^{*}$ & -0.001$^{**}$ & -0.001 & -0.004$^{**}$ \\
& -- & -- & (0.000) & (0.000) & (0.001) & (0.001) \\
collab. count & -- & -- & 0.227$^{***}$ & 0.197$^{***}$ & 0.162$^{***}$ & 0.126$^{***}$ \\
& -- & -- & (0.010) & (0.009) & (0.015) & (0.017) \\
familiarity$\times$status & -- & -- & -- & -- & -0.000 & 0.006$^{***}$ \\
& -- & -- & -- & -- & (0.001) & (0.001) \\
familiarity$\times$count & -- & -- & -- & -- & 0.088$^{***}$ & 0.194$^{***}$ \\
& -- & -- & -- & -- & (0.014) & (0.026) \\
status$\times$count & -- & -- & -- & -- & 0.000 & 0.001$^{**}$ \\
& -- & -- & -- & -- & (0.000) & (0.000) \\
familiarity$\times$status$\times$count & -- & -- & -- & -- & -- & -0.003$^{***}$ \\
& -- & -- & -- & -- & -- & (0.001) \\
\hline
two-way fixed effects & Y & Y & Y & Y & Y & Y \\
Observations & 11,510 & 16,877 & 16,615 & 11,359 & 11,359 & 11,359 \\
N. of groups & 2,817 & 5,360 & 5,309 & 2,804 & 2,804 & 2,804 \\
$R^2$ & 0.287 & 0.156 & 0.609 & 0.671 & 0.689 & 0.695 \\
Residual Std. Error & 0.743 & 0.497 & 0.987 & 1.141 & 1.156 & 1.161 \\
F Statistic & 1747.213$^{***}$ & 2119.285$^{***}$ & 5849.378$^{***}$ & 2904.998$^{***}$ & 2099.285$^{***}$ & 1941.321$^{***}$ \\
\hline
\multicolumn{7}{r}{\textit{Note:} $^{*}$p$<$0.05; $^{**}$p$<$0.01; $^{***}$p$<$0.001} \\
\end{longtable}
\end{center}
\end{landscape}

Research persistence shows contrasting \likai{associations with} productivity and impact. It is positively associated with productivity both before and after the paradigm shift - researchers who remained committed to their established topics or methods tended to publish more papers. \likai{The lower productivity of adaptive researchers} could be due to several factors: they may have begun submitting to ICLR, a conference focused on deep learning, or needed time to regain their previous productivity levels while learning new methods and adjusting to the emerging paradigm (see Appendix \ref{appendix:alternatives} for more discussion). On the other hand, the interaction term between persistence and the post-2013 period reveals a significant negative relationship with impact, indicating that \likai{maintaining early research directions was associated with less influential work} after the deep learning boom. \likai{In the era following the advent of deep learning, given identical coauthor networks and prior performance, a 0.1 increase in similarity to prior research is associated with a 1.77 percentile-point decrease in impact; equivalently, a one-standard-deviation increase in persistence is associated with a decrease of approximately 2.7 percentile points. For productivity (Table~\ref{prod_table}), a 0.1 increase in persistence is associated with about 0.28 more papers per year after 2013.} \revtwo{Departing from prior work after 2013 was thus associated with higher citation impact but lower output, a trade-off we examine across the full range of persistence in Section~\ref{5.2.3}.}

\likai{The timing of the penalty is informative. In an event-study version of the model that replaces the pooled post-2013 interaction with year-specific interactions (base year 2012; Appendix~\ref{causal}), the pre-2013 coefficients are jointly insignificant (impact: $p=.347$; productivity: $p=.505$), so we find no evidence that high- and low-persistence scientists were already diverging before the shift. The impact penalty is near zero in 2013--2015, turns negative thereafter, and reaches $-0.276$ ($p=.014$) by 2022, indicating that it accumulated over time rather than appearing at once. The difference between the mean post- and pre-period coefficients ($-0.184$) is consistent with the pooled estimate in Table~\ref{impact_table} ($-0.177$). The positive association between persistence and productivity is unaffected. The results are also robust to the alternative persistence measures in Appendix~\ref{persistence_m}: per standard deviation of persistence, impact falls by 2.27 percentile points under the five-year-window measure and by 2.59 under the LDA topic-level measure, while productivity rises in both cases.}

\begin{figure}[htbp]
    \centering
   \includegraphics[width=0.85\textwidth]{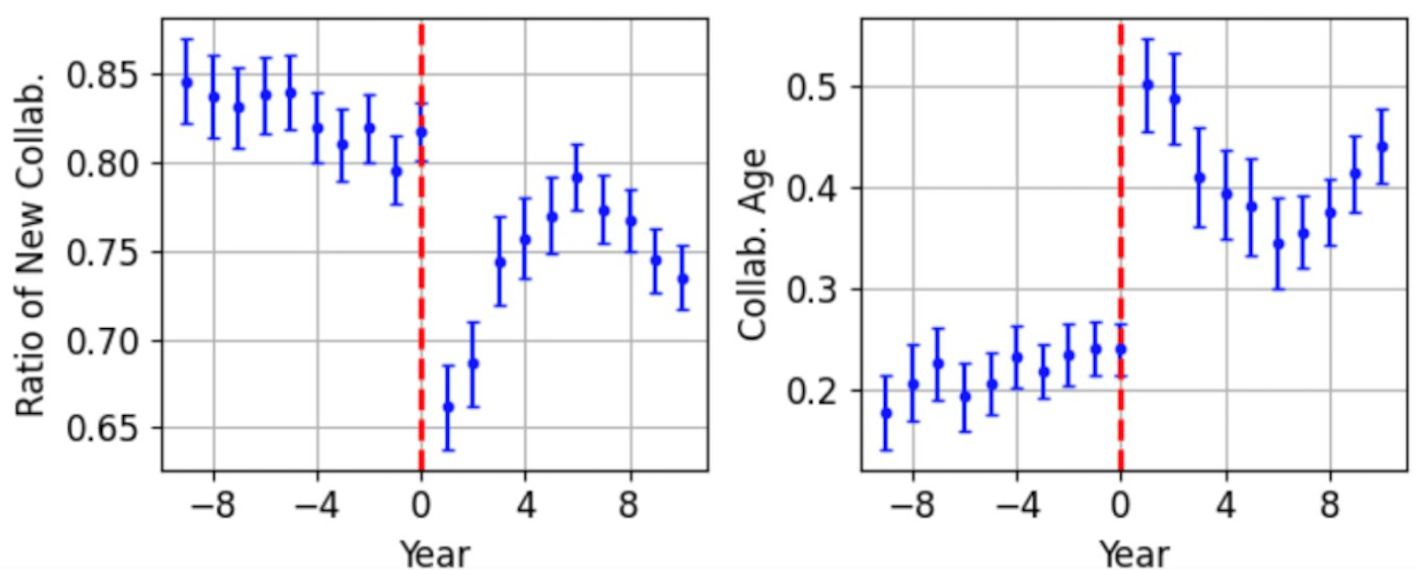} % full width
    \footnotesize
     \caption{Teams adapt through fresh collaborations. \likai{Left: share of new collaborators, i.e., the percentage of a publishing scientist's coauthors in a given year who had never coauthored an ICML/NeurIPS paper with them before. Right: team age (collaboration familiarity), i.e., the average number of prior ICML/NeurIPS papers the scientist had coauthored with that year's coauthors, tracked back to the inception of both conferences. Year 1 corresponds to 2013.} Error bars represent 95\% confidence intervals.}
\label{fig:age2}
\end{figure}

Collaborator familiarity, count, and status exhibit mixed \likai{associations with} productivity and impact. Having more collaborators \likai{is associated with higher} productivity and impact. Collaborating with high-status partners \likai{is associated with significantly higher impact}. However, this advantage comes with a tradeoff, as such collaborations tend to \likai{be associated with lower} productivity, potentially due to the time and resources required to maintain these high-status relationships and high-status authors’ high standard of publishing. Teams with older collaborations exhibit lower productivity and impact levels\likai{, consistent with longer-standing teams adapting less successfully} to the new paradigm. We further examined the "age" of these scientific teams as shown in Figure \ref{fig:age2}, defined as the average number of ICML/NeurIPS papers the focal author had previously collaborated on with team members that year (right), and the proportion of new collaborations, defined as the percentage of team members who had never co-authored an ICML/NeurIPS paper with the focal author before (left). We tracked their collaborations back to the initial year of these conferences. We found that right after the shock, published teams tended to be more experienced — presumably those at the center of deep learning prior to the shock - and they continued to adapt through collaborating with newcomers in the years following 2013. This is evidenced by a decreasing average team age and an increasing share of new collaborators, both of which \likai{coincided with} the continued success. Notably, these new collaborations tended to persist over time, as teams gradually became old again in subsequent years. These effects are robust within individual conferences; see Appendix \ref{appendix:within}.

%This is also validated in Figure \ref{fig:age2} – teams that stay after 2013 consistently boost productivity through new collaborations. 

\likai{We also examine whether natural aging contributes to the redistribution of influence.} Over the (brief) decade of post-deep-learning we tracked, it is possible that some scientists’ drive increased (e.g., those in their career growth phase), while others’ drive declined (e.g., those who were already well-established). In Appendix \ref{appendix:age_appendix}, we track productivity and impact by academic age (the years since their first publication), controlling for author fixed effects, across the cohort of scientists and find no strong evidence that impact or productivity decreases with age. In fact, while there is no meaningful \likai{association with} impact, productivity actually increases with age.

\subsubsection{Who will be more adaptive?}
\label{6.2.2}
Who is more likely to adapt? Adaptivity is \likai{closely associated with} both impact and productivity in the post-deep learning era. If previously high-status scientists become less adaptive, they will lose status which could result in a redistribution of influence as observed in Section \ref{6.2.1}. This section explores this critical issue by examining the features of authors and their adaptiveness. We test it using the following regression:

\begin{align*}
\textit{Research Persistence}_{i,t} =\; 
& \beta_1 \times \textit{Prev. Impact}_{i,t} + \beta_2 \times \textit{Prev. Productivity}_{i,t} \\
& + \beta_3 \times \textit{Collaborator Familiarity}_{i,t} + \beta_4 \times \textit{Collaborator Count}_{i,t} \\
& + \beta_5 \times \textit{Collaborator Status}_{i,t} \\
& + \textit{Potential Interactions (Collab. Familiarity}_{i,t}, \textit{ Count}_{i,t}, \textit{ Status}_{i,t}) \\
& + \alpha_i + \gamma_t + \varepsilon_{i,t}
\end{align*}

\begin{landscape}
\begin{center}
\footnotesize
\begin{longtable}{@{\extracolsep{5pt}}lcccc}
\caption{Regression Results: Persistence as Dependent Variable} \label{persist_table}\\
\toprule
& \multicolumn{4}{c}{\textit{Dependent variable: persistence}} \\
\cmidrule(lr){2-5}
& (1) & (2) & (3) & (4) \\
\midrule
\endfirsthead

\multicolumn{5}{c}%
{\tablename\ \thetable\ -- \textit{Continued from previous page}} \\
\toprule
& \multicolumn{4}{c}{\textit{Dependent variable: persistence}} \\
\cmidrule(lr){2-5}
& (1) & (2) & (3) & (4) \\
\midrule
\endhead

\midrule \multicolumn{5}{r}{\textit{Continued on next page}} \\
\endfoot

\bottomrule
\endlastfoot

prev. productivity & 4.040$^{***}$ & -- & -- & 3.236$^{***}$ \\
& (0.150) & -- & -- & (0.127) \\
prev. impact & -- & 0.019$^{**}$ & -- & 0.015$^{**}$ \\
& -- & (0.006) & -- & (0.006) \\
collab. familiarity & -- & -- & 1.828$^{*}$ & 1.995$^{**}$ \\
& -- & -- & (0.772) & (0.738) \\
collab. status & -- & -- & 0.008 & 0.008 \\
& -- & -- & (0.012) & (0.011) \\
collab. count & -- & -- & 0.618$^{***}$ & 0.635$^{***}$ \\
& -- & -- & (0.107) & (0.101) \\
familiarity\_status & -- & -- & -0.002 & -0.002 \\
& -- & -- & (0.014) & (0.014) \\
familiarity\_count & -- & -- & 0.833$^{***}$ & 0.503$^{*}$ \\
& -- & -- & (0.213) & (0.207) \\
count\_status & -- & -- & 0.003 & -0.000 \\
& -- & -- & (0.003) & (0.003) \\
familiarity\_count\_status & -- & -- & -0.005 & -0.011$^{*}$ \\
& -- & -- & (0.004) & (0.004) \\
\midrule
two-way fixed effects & Y & Y & Y & Y \\
Observations & 11,510 & 11,319 & 11,359 & 11,170 \\
N. of groups & 2,817 & 2,787 & 2,804 & 2,772 \\
$R^2$ & 0.234 & 0.001 & 0.193 & 0.305 \\
Residual Std. Error & 5.825 & 0.444 & 5.279 & 6.636 \\
F Statistic & 2,652.971$^{***}$ & 11.620$^{***}$ & 290.508$^{***}$ & 408.376$^{***}$ \\
& (df=2,836; 8,674) & (df=2,806; 8,513) & (df=2,829; 8,530) & (df=2,799; 8,371) \\
\midrule
\multicolumn{5}{r}{\textit{Note:} $^{*}$p$<$0.05; $^{**}$p$<$0.01; $^{***}$p$<$0.001} \\
\end{longtable}
\end{center}
\end{landscape}

Our analysis reveals that a scientist’s past influence and productivity are \likai{significantly associated with} the trajectory of their research evolution. Prolific scholars tend to be less inclined to shift their research paradigms. Older and larger teams are less likely to adapt. These findings suggest that in a dynamic scientific environment, researchers who adopt high-yield strategies — particularly those characterized by visible and immediate signals such as productivity or collaboration with established, large teams — become increasingly reluctant to abandon them, while such persistence \likai{is associated with lower subsequent impact}. \revtwo{Because persistence continues to sustain output, this reluctance need not be irrational: scientists with the most output at stake may stand to lose the most from adapting.}

\begin{figure}[htbp]
    \centering
    \includegraphics[width=1\textwidth]{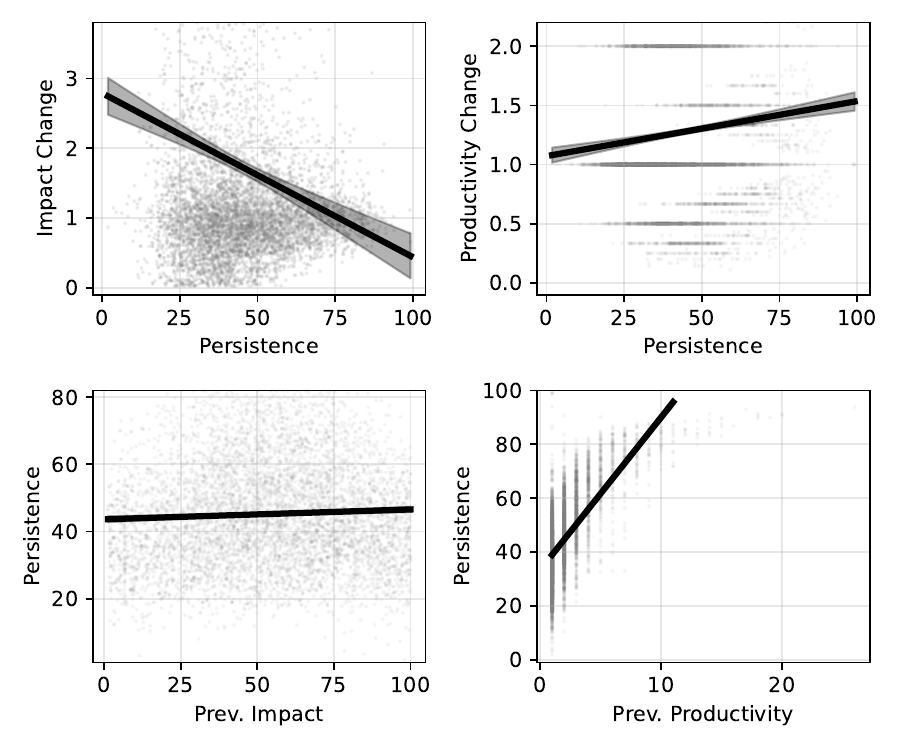} % full width
    \footnotesize
    \caption{The relationship between persistence, status, and status change using linear fits. \likai{The fits are descriptive; formal tests of functional form, with controls and two-way fixed effects, are reported in Appendix~\ref{causal}.} Error bands represent 95\% confidence intervals.}
    \label{fig:adapt}
\end{figure}
We present the relationship between status and persistence more intuitively. To measure change, we calculated the ratios of productivity and impact relative to the previous year. Figure \ref{fig:adapt} illustrates how impact and productivity evolve with increasing persistence (top row) and how persistence varies with previous productivity and impact (bottom row). A clear trend emerges: as persistence increases, impact tends to decrease while productivity increases. \likai{Additionally, as previous impact increases, persistence shows a weak upward trend, whereas previous productivity is strongly correlated with increased persistence (bottom row).}

\subsubsection{\likai{Where the field sits versus where \revtwo{per-author benefit} is highest}}
\label{5.2.3}
\likai{Section~\ref{section3} noted that prior work suggests moderate adaptation might be a natural choice for scientists when they face the paradigm shift. We test this formally by replacing the linear persistence term with ten persistence bins (0--10, \ldots, 90--100; reference bin 30--40), while keeping all controls and two-way fixed effects (Appendix~\ref{causal}.4). At the \textit{collective level} of the field, we indeed find that moderate persistence dominates. Among post-2012 author-years, 59.7\% fall in the 25--50 band, which accounts for 42.2\% of total output, 41.7\% of impact-weighted output (annual papers multiplied by their citation percentile), and 68.1\% of all top-decile author-years. An author-clustered bootstrap over 1,000 replications places the maximum in this band with probability 1.00 for total output, 0.998 for impact-weighted output, and 1.00 for top-decile author-years. However, at the \textit{individual level}, we do not find an interior optimum in persistence. Across ten bins and five outcomes, impact is strictly decreasing in persistence, whereas productivity and impact-weighted output are strictly increasing, with no change in the sign of the slope. Because a monotone mean does not preclude a hump in the upper tail, we separately test the probability of reaching the top 10\% and top 5\% of impact. Both peak at the low end of persistence, and Lind--Mehlum tests \citep{lind2010with} support neither an inverted U ($p=1.00$ for both tail outcomes) nor a U shape ($p=.90$ and $p=.57$). The shape conclusion also holds on the 20--80 support, which contains 93.5\% of the sample.}

\likai{Both facts hold simultaneously because the aggregate peak is a headcount effect: the 25--50 band contributes the most because it contains the most people, not because each individual performs best, and its share of output is lower than its share of author-years. A steadily rising per-person curve and a hump-shaped total are compatible whenever the population thins out sufficiently quickly at the high end, as it does here. The 25--50 band is therefore the \textit{modal} range of adaptation. \revtwo{Note, however, that no single outcome peaks in this band either: impact is highest at low persistence, whereas productivity and impact-weighted output are highest at high persistence. A scientist who values both output and impact could therefore reasonably settle between the two extremes, and most scientists indeed do so. We discuss this in more detail in Section~\ref{section7}.}}

\subsection{\texorpdfstring{\likai{Alternative Explanations and Complementary Analyses}}{Alternative Explanations and Complementary Analyses}}
\label{sec:alt}

\paragraph{\likai{Survivorship bias.}}
\likai{Our panel is unbalanced: author-year observations peak at 1,398 in 2012, fall to 791 in 2013 and 588 in 2015, and recover gradually thereafter (Appendix~\ref{survivorship}). If the scientists who left ICML and NeurIPS were disproportionately flexible but unsuccessful, while those who stayed were disproportionately persistent but successful, survivorship could manufacture or mask the rigidity penalty. We examine this in three steps (Appendix~\ref{survivorship}). First, an exit model on pre-2013 characteristics shows that attrition is downward. Conditional on the other covariates, a one-standard-deviation increase in prior productivity multiplies the odds of exit by 0.45 ($p<.001$), with prior impact (OR $=0.84$, $p=.001$) and academic age (OR $=0.89$, $p=.026$) pointing the same way, whereas pre-2013 persistence has no independent association with exit ($p=.159$). The panel thus loses low-output, low-impact, junior scientists rather than selecting on persistence. Second, citations to the cohort's pre-2013 papers, counted across all of OpenAlex rather than only within the conferences, are lower for leavers than for stayers in every post-2013 year. Third, we bound the estimate by returning exiters to the panel. Under a neutral rule in which exiters carry their own pre-period level, the penalty is slightly stronger ($-0.193$ versus $-0.177$, $p<.001$).}

\paragraph{\likai{Topic popularity: the field's behavior or the scientist's?}}
\likai{A scientist who keeps working on a declining topic will register high persistence while their citations fall as the field moves on, without persistence itself playing any role. Because persistence is defined relative to the scientist's own prior work rather than to the field, it does not itself decline as the field moves; the concern is that it may co-move with topic decline. Relatedly, our outcome is defined within these conferences, so the penalty might merely reflect these venues' growing preference for deep learning topics. We address these concerns in Appendix~\ref{topic_nomal}, using a 30-topic LDA model fit on the corpus. First, we construct \textit{topic momentum}, the year-over-year change in the share of the topics that authors actually worked on for each author-year, and add it to the main regression. Topic momentum is itself a strong predictor of impact ($+0.67$, $p=.001$), so the mechanical channel is real, but the persistence coefficient for impact barely moves ($-0.154$ to $-0.152$, $p=.003$). Adding author-year dominant-topic fixed effects moves the persistence coefficient for impact to $-0.167$ ($p<.001$). The persistence coefficient and significance for productivity are unchanged. Second, measuring impact as a citation percentile among papers on the same topic in the same year leaves the persistence coefficient to be $-0.124$ ($p=.011$), roughly a 20\% attenuation. Topic popularity thus accounts for about a fifth of the penalty, not the whole of it: persistent scientists are cited less than other scientists working on the same topic, in the same year, at the same venue. We nonetheless recognize that the component of impact reflecting venue selection cannot be fully separated from differences in research quality, a point we return to in Section~\ref{section7}.}

\paragraph{\likai{Complementary causal analysis.}}
\likai{We complement the panel model with two designs that relax different assumptions (Appendix~\ref{causal}); neither delivers an identified causal effect, and we label them accordingly. The first is a \textit{quasi-experimental heterogeneity analysis} of pre-shock proximity to deep learning, measured as the SPECTER2 cosine similarity between each author's 2003--2012 papers and a deep-learning centroid. Comparing authors locally around the median, scientists whose pre-shock work was already closer to deep learning had significantly higher post-2013 productivity ($+0.157$, $p=.019$) and marginally higher persistence ($+0.012$, $p=.062$): they were already working on the winning topics and had less need to move. Because the running variable is smoothly distributed with no natural discontinuity, a regression-discontinuity design is not available, and we do not present this analysis as one. The second is a \textit{selection-on-observables} propensity-score matching analysis. Among post-2013 survivors, we split scientists at the median of their average post-2013 persistence, estimate a propensity score from seven pre-2013 characteristics (productivity, impact, coauthor status, number of collaborators, collaboration age, academic age, and proximity to deep learning (as in the quasi-experimental heterogeneity analysis)), and match one-to-one, yielding 926 pairs with all standardized mean differences at or below 0.12. In the matched sample, persistent scientists published about 0.84 more papers per year after 2013 ($p<.001$); their impact was about one percentile point lower, in the same direction as the main result but not statistically significant ($p=.34$). The loss of power is expected, because this design assigns each scientist a single label for the whole post-2013 decade, whereas the panel model uses year-to-year variation.}

\subsection{Summary of Results}
\label{6.3}

\begin{figure}[htbp]
    \centering
    \includegraphics[width=0.8\textwidth]{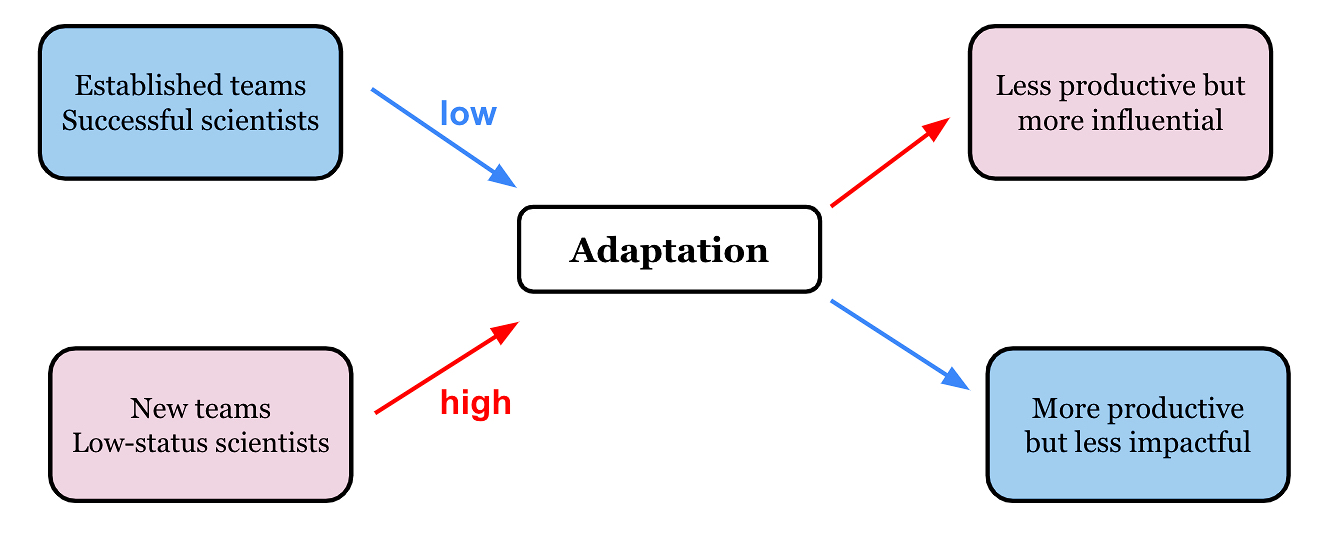} % full width
    \footnotesize
    \caption{Summary of key results.}
    \label{fig:summary}
\end{figure}

\likai{Overall, persistence, measured as the semantic similarity of a scientist's new work to their own prior work, is positively associated with productivity but, after 2013, negatively associated with scientific impact as measured by citation percentile rank. Scientists who were previously successful or affiliated with established teams adapted more slowly, showing higher persistence. Prior impact is negatively associated with subsequent impact; productivity shows a similar but statistically insignificant pattern once interactions among collaboration-network features are controlled\footnote{This pattern likely reflects, at least in part, regression to the mean: extreme prior values, measured with noise, tend to revert toward the mean in subsequent periods. In addition, including a lagged dependent variable alongside author fixed effects mechanically biases its coefficient downward \citep{nickell1981biases}. Because prior impact enters only as a control, and the post-2013 interaction term is similar with and without it (Table~1, Model~1 vs.\ Models~4--6), these biases are unlikely to drive our estimates of the persistence terms.}. Together, these results (Figure~\ref{fig:summary}) point to a tension between persistence and adaptation during the deep learning revolution. Researchers who maintained continuity with their prior work tended to be more productive, both before and after 2013. After the shift, however, greater persistence was associated with lower citation impact, a penalty that accumulated over the following decade (Appendix \ref{app:event}), whereas scientists who moved further from their prior work tended to produce more influential contributions, at the cost of lower output. Scientists who were previously successful or embedded in long-standing teams were slower to adapt, consistent with the shift opening room for those less invested in the old paradigm. Yet most scientists clustered in a band of moderate persistence, even though per-author impact kept rising with greater departure from prior work\revtwo{, a position between the productivity gains of persistence and the impact gains of adaptation}.}

Interestingly, our results suggest that the motivations and capacities for adaptation and persistence vary across different strata of the scientific community. \likai{Scientists with less prior success, less constrained by entrenched positions \revtwo{and with less output at stake}, may feel a heightened urgency to explore new directions during periods of disruption, which may allow them to capitalize on emerging opportunities and challenge existing power dynamics.} This interplay of disruption, status, and adaptability underscores the complexity of scientific evolution, where both the persistence of existing paradigms and the emergence of novel directions coexist and compete.

\section{\texorpdfstring{\likai{Discussion}}{Discussion}}
\label{section7}
This study examines the dynamics of scientific progress during a major paradigm shift in the field of machine learning, focusing on how researchers’ strategies between persistence and adaptability \likai{relate to} future success. While persistence is positively associated with productivity, during the post-2013 deep learning era, it is negatively associated with impact. This finding suggests that rigid adherence to earlier research directions, especially during a major paradigm shift, \likai{is associated with a reduced capacity} to produce influential work in rapidly evolving fields\revtwo{, even as it continues to sustain output}. \likai{The event-study estimates (Appendix \ref{app:event}) indicate that this penalty accumulated gradually over the decade rather than appearing at once, and it is robust to alternative persistence measures, controls for topic popularity, different identification strategies, and neutral survivorship bounds (Appendix \ref{persistence_m} to \ref{causal}).}

\likai{A second finding concerns the gap between \revtwo{per-author benefits} and the distribution of scientists. At the individual level, we do not claim that a particular degree of overlap with prior work is optimal for any given researcher. At the level of the field, however, most scientists, and \revtwo{the largest share} of the field's (impact-weighted) output, are concentrated in a band of moderate persistence (25--50\% overlap with prior work), where per-author output lies below the field average. Compared to empirical evidence that adaptive researchers are often penalized in the scientific system \citep{leahey2017prominent, schilling2011recombinant, wagner2011approaches, hill2025pivot}, our results suggest that during a paradigm-shifting event the relationship between adaptation and impact is more contextual and dynamic: as the community changes, so do the criteria of research evaluation. Why, then, do scientists cluster where they do? \revtwo{Our results are consistent with two families of explanations, which our data cannot fully separate. The first is a trade-off: because persistence is associated with higher productivity and adaptation with higher impact, a scientist who values both may rationally settle in between. An interior choice of this kind requires that scientists value the two outcomes with diminishing returns rather than simply in combination: impact-weighted output, which multiplies them, still rises monotonically with persistence (Table~\ref{tab:bins}), so a scientist maximizing either product alone would not stop in the moderate band. The second is friction: scientists may wish to move further but face barriers that make large departures costly, including career risk, the cost and time lag of retooling, amplified by the burden of knowledge \citep{jones2009burden}, and funding processes that reward incremental extensions of an established line of work \citep{boudreau2012novelty, ayoubi2021does, shin2022scientific}. The two are not mutually exclusive, and both are institutional: the value scientists place on productivity or impact is itself shaped by how institutions evaluate them. A third possibility is selection on heterogeneous returns. The scientists who moved furthest may have been those with the most to gain, for example because their skills transferred well to deep learning, so the impact they achieved need not be available to those who stayed in the moderate band. Distinguishing these explanations, for instance by comparing scientists under different topics or at career stages that face different productivity pressures, is an important task for future work.}}

The interplay between author status and adaptability further highlights the redistributive effects of disruption. \likai{Scientists with less prior success, often constrained by limited resources and recognition, appear more likely to adapt during periods of disruption, which may allow them to reposition themselves within the scientific hierarchy. Conversely, previously successful scientists and large, long-standing teams, while better resourced, face stronger incentives to maintain their existing research lines\revtwo{, which continue to sustain their output,} and adapt more slowly. This dynamic is associated with a redistribution of impact, in which disruption opens opportunities for some while reinforcing barriers for others. It complements the evidence of \citet{azoulay2019does}: where they show that the exit of dominant incumbents opens a field to outsiders; our results suggest that a similar opportunity structure can arise from the \textit{persistent presence} of incumbents whose expertise the shift has devalued. Such situations challenge traditional hierarchies, and those who adapt, often through fresh collaborations, tend to \revtwo{gain in impact, if at some cost to output}. Disruption can thus act both as a destabilizing force and as a catalyst for innovation, pushing scientists to confront the limits of their existing approaches.}

Beyond findings at the micro-level, we also observe how ICML and NeurIPS conferences adapted to the deep learning paradigm, increasingly aligning with the cutting-edge topics emphasized at ICLR. Trends in topic importance demonstrate a convergence among these top conferences, possibly achieved by active adaptation through peer review and selection processes, as well as passive adaptation through cross-conference submissions. This adaptation highlights how leading scientific venues respond to disruptive changes to maintain their relevance and influence. The alignment of ICML and NeurIPS with ICLR reflects broader institutional flexibility to incorporate new paradigms, a critical factor for sustaining their roles as flagship venues. \likai{This venue-level adaptation is part of the phenomenon we study rather than a nuisance to it: the rigidity penalty is jointly produced by shifting scientific evaluation and shifting venue selection. Our design can separate the topic-popularity component, which accounts for roughly a fifth of the penalty, but it cannot separate reviewers' preferences from genuine differences in quality. That distinction is inherently difficult to sustain in this context. ICML and NeurIPS are staffed by leading scientists in the field, whose editorial judgments reflect deep expertise and broad consensus about what constitutes valuable research and fruitful directions; their collective preference is therefore not arbitrary, but a meaningful signal of research quality within the field's prevailing paradigm, even though it cannot be cleanly separated from topic fashion.}

Our findings have important implications for individual researchers, institutions, funding agencies, and future research evaluation exercises. For researchers, \likai{our results indicate that the returns to persistence can shift sharply during periods of disruption \revtwo{and that the choice involves a trade-off between output and impact}, although they do not identify an optimal degree of adaptation for any individual.} Institutions and funding agencies should recognize the value of both research strategies, providing support for researchers to explore new directions while maintaining their foundational expertise\likai{; because publication venues and funding bodies move at different speeds, uncoordinated signals can leave scientists penalized by one system for what the other rewards}. \revtwo{If scientists' clustering in the moderate band reflects, in part, evaluation systems that reward publication counts, then such systems may themselves discourage the adaptation that a paradigm shift rewards.} Additionally, our findings raise a question about the nature of research evaluation: research articles often lack visible signals for quality, and researchers frequently rely on heuristics such as citation counts or authors' status \citep{azoulay2014matthew, bao2024simulation} when they cite. As the trendiness of a topic is another heuristic that could boost citations, its relationship with the quality of work is often highly blurred during a major shift in the research paradigm. \likai{Our impact measure should accordingly not be read as universal scientific impact: it captures recognition within a specific set of selective venues during a technological shock, and our sample may not fully capture the recognition eventually received by researchers who pursued deep, non-conformist work outside the dominant paradigm.} Effectively evaluating the quality of publications and evaluating the appropriate emphasis on productivity versus impact in evolving academic landscapes remain a critical challenge for all scientific communities.

\likai{Despite our timely contribution to understanding the persistence of scientific careers, our study has several limitations that readers should note and that future work should address. The first concerns what our data can observe. Because we analyze published papers, we capture only part of the adaptation process: adaptive mechanisms may also operate during peer review, as authors revise in response to referees. Analyzing submitted manuscripts alongside their published versions would offer a more comprehensive view of how papers adapt and how that adaptation shapes their success. Such an analysis is currently constrained by data availability, since review records exist only for the most recent years, but the continued development of open peer review infrastructures may make this research feasible in the future. This blind spot is consequential because review may itself be part of the mechanism we document: journal peer review favors novel manuscripts, and editors select on novelty even conditional on reviewer scores \citep{teplitskiy2022novel}. To the extent that gatekeeping operates this way, some portion of the persistence penalty may reflect selection at the review stage. A second measurement limitation is our reliance on textual similarity. Future work could move beyond embedding-based textual similarity by tracing how novel ideas and methods propagate through co-authorship and citation networks. Such an approach would provide more direct evidence of how weak ties to prior expertise operate in practice. During a technological shock, citation impact at selective venues is an imperfect proxy for the long-run value of non-conformist work. A more nuanced measurement approach would also be valuable, one that distinguishes a topic's popularity with the program committees of leading conferences from its genuine intellectual merit. The two overlap heavily but are not identical, and the difference matters: without drawing it, the institution risks rewarding, and implicitly endorsing, the chasing of hot topics over the pursuit of durable contributions.}

\likai{The third set of limitations concerns the scope of our conclusions, which are best situated within the broader literature on scientific conservatism and the "essential tension" between tradition and innovation in scientific work \citep{kuhn1977essential}. Persistence and adaptation are not simply competing strategies with one dominant winner; they reflect an enduring and often productive tension, in which disciplinary commitment sustains the cumulative, non-conformist inquiry from which foundational breakthroughs eventually emerge, while responsiveness to new paradigms allows a field to absorb genuine advances. Read against this backdrop, our results do not resolve this tension in favor of adaptation as a general prescription. Rather, they document how the balance tipped sharply during one specific, high-velocity episode. The deep learning revolution is atypical in several respects: it was industry-accelerated, adopted at unusual speed, catalyzed several further breakthroughs that continue to shape the field, and unfolded within a publication culture defined by fast conference cycles and rapid citation turnover \citep{park2023papers, mukhamediev2021classical, singh2023forgotten}. These features likely amplify the rigidity penalty we document beyond what would be observed in fields where paradigm shifts unfold more gradually and the costs of persistence accumulate more slowly. Our evidence therefore rests on a single, unusually large paradigm shift, which limits its representativeness for disciplines organized around journals and multi-year review cycles. We thus read our results as characterizing the persistence-adaptation trade-off as an outcome of conditions particular to this shift, rather than as a universal law of scientific careers or an endorsement of trend-chasing over deep, foundational inquiry. Coherence with a scientist's prior work and alignment with prevailing interests are weighted very differently by evaluators across fields, indicating that the returns to persistence also depend on the evaluating community \likai{\citep{ayoubi2026research}}. Comparing paradigm shifts across fields with different publication cultures, paces of change, and institutional structures would help establish the scope conditions under which the persistence paradox generalizes, and would allow a more direct test of how the essential tension resolves differently across settings.}

\vspace{\baselineskip}

\textbf{Acknowledge}: We thank Hongyuan Xia, Yufeng Xia, Nadav Kunievsky, James A. Evans, participants at the ASIS\&T MET-STI workshop, and members of the UChicago Knowledge Lab for valuable discussions. This research was supported in part by the University of Chicago Research Computing Center. All errors are our own.

\newpage

\bibliographystyle{chicago}
\bibliography{Bibliography-MM-MC}

\newpage
\appendix

% Number appendix tables/figures by section (A1, B2, ...), so they can be
% cited unambiguously in the response letter.
\counterwithin{table}{section}
\counterwithin{figure}{section}
\renewcommand{\thetable}{\thesection\arabic{table}}
\renewcommand{\thefigure}{\thesection\arabic{figure}}

\appendix
\section*{Appendix}   % 如果你原来就有这个标题，保留即可

\startcontents[appendix]
\printcontents[appendix]{}{1}{\subsection*{Contents}}

\newpage
\section{\texorpdfstring{Robustness of persistence measures}{Robustness of persistence measures}}
\label{persistence_m}

\likai{Our main persistence measure compares each scientist's papers in a given year with all of their prior ICML/NeurIPS papers, using bootstrapped SPECTER2 embedding similarity on titles and abstracts. Because different text-representation algorithms can yield substantially different similarity scores for the same documents \citep{tshitoyan2019unsupervised}, and because two design choices in the main measure could in principle drive the results, we re-estimate the main model under two alternative operationalizations (Table~\ref{tab:persistence_alt}).}

\likai{
\begin{table}[htbp]
\centering
\footnotesize
\caption{Main model under alternative persistence measures, standardized effect per +1 SD of persistence.}
\label{tab:persistence_alt}
\begin{tabular}{lccc}
\toprule
Persistence measure & Impact, per +1 SD & Productivity, per +1 SD & Author-years \\
\midrule
Five-year window & $-2.27^{**}$ & $0.249^{***}$ & 10,576 \\
 & (0.76) & (0.035) & \\
LDA topic distribution ($k=30$) & $-2.59^{***}$ & $0.083^{***}$ & 11,649 \\
 & (0.70) & (0.025) & \\
\bottomrule
\end{tabular}
\par\smallskip
\begin{minipage}{0.9\textwidth}\footnotesize
\textit{Notes.} Two-way (author and year) fixed effects, author-clustered standard errors, and controls are the same as the main specification. Standard errors in parentheses. $^{*}p<.05$, $^{**}p<.01$, $^{***}p<.001$.
\end{minipage}
\end{table}
}

\likai{\textbf{Five-year window.} The main measure uses the full prior corpus because the construct of interest is distance from \textit{accumulated} expertise, which is what a paradigm shift devalues; a rolling window would treat a scientist who abandoned a topic fifteen years ago as identical to one who never worked on it. This choice is nevertheless consequential for long-career authors, whose prior corpora may span several shifts in direction. We therefore construct a version of the measure in which the prior corpus is restricted to the five years preceding the focal year. The two measures correlate at $r=0.87$. Per standard deviation of persistence, impact falls by 2.27 percentile points ($p=.003$) and productivity rises by 0.249 papers per year ($p<.001$), compared with approximately 2.7 percentile points per standard deviation for impact in the main specification. Author fixed effects additionally absorb time-invariant differences in career length.}

\likai{\textbf{LDA topic distribution.} The main measure equalizes corpus sizes by bootstrap subsampling, because embedding similarity is mechanically depressed when a small set of documents is compared with a much larger one. In most author-years the focal-year corpus is the smaller of the two, so the procedure subsamples the prior corpus. To verify that the results do not depend on this step, we construct a measure that involves no size matching at all. We fit a 30-topic Latent Dirichlet Allocation (LDA) model on the full ICLR-inclusive corpus, represent each author-year by the aggregate topic distribution of that year's papers and by the aggregate topic distribution of all prior papers, and define topical persistence as the cosine similarity between the two distributions. This measure reproduces both results: impact falls by 2.59 percentile points per standard deviation ($p<.001$) and productivity rises by 0.083 papers per year ($p<.001$). These results indicate that the rigidity penalty does not depend on the choice of text representation, the length of the reference window, or the size-matching procedure.}

\newpage
\section{\texorpdfstring{Survivorship bias}{Survivorship bias}}
\label{survivorship}

\likai{Our main estimates use the unbalanced panel of cohort scientists observed in each year in which they publish at ICML or NeurIPS. Scientists who stop publishing at these venues after 2013 drop out of the post-period. This raises the concern that the sample of ``survivors'' is selected: if rigidity led scientists to exit these venues entirely, the true cost of rigidity would not be captured, and if survivors disproportionately include scientists who were successful regardless of how they adapted, the estimated returns to adaptation could be biased. We examine the attrition structure, model who exits, and bound the estimate.}

\likai{\textbf{Attrition structure.} Table~\ref{tab:obs_by_year} reports author-year observations by year. Observations rise steadily to a peak of 1,398 in 2012, fall sharply to 791 in 2013 and 588 in 2015, and then recover gradually, reaching 841 by 2022, reflecting the time and effort needed for adaptation.}

\likai{
\begin{table}[htbp]
\centering
\footnotesize
\caption{Panel observations by year.}
\label{tab:obs_by_year}
\begin{tabular}{cc@{\hspace{2.5em}}cc}
\toprule
Year & Author-year observations & Year & Author-year observations \\
\midrule
2003 & 688 & 2013 & 791 \\
2004 & 728 & 2014 & 772 \\
2005 & 804 & 2015 & 588 \\
2006 & 797 & 2016 & 688 \\
2007 & 850 & 2017 & 663 \\
2008 & 949 & 2018 & 727 \\
2009 & 1,007 & 2019 & 770 \\
2010 & 1,043 & 2020 & 819 \\
2011 & 1,089 & 2021 & 865 \\
2012 & 1,398 & 2022 & 841 \\
\bottomrule
\end{tabular}
\end{table}
}

\likai{\textbf{Who exits.} Table~\ref{tab:exit} models exit as a function of pre-2013 characteristics in a single multivariate logit. Conditional on the other covariates, prior productivity is the dominant predictor: a one-standard-deviation increase multiplies the odds of exit by 0.45 ($p<.001$). Prior impact (OR $=0.844$, $p=.001$) and academic age (OR $=0.888$, $p=.026$) point the same way. Attrition is therefore downward: the panel loses low-output, low-impact, junior scientists. Once productivity is controlled, pre-2013 persistence has no independent association with exit ($p=.159$).}

\likai{
\begin{table}[htbp]
\centering
\footnotesize
\caption{Does attrition select on persistence, or on productivity and status?}
\label{tab:exit}
\begin{tabular}{lccccc}
\toprule
Predictor (z-scored) & Exited mean & Stayed mean & Odds ratio & $\beta$ & $p$ \\
\midrule
Pre-2013 persistence & 0.344 & 0.363 & 1.086 & 0.082 (0.058) & .159 \\
Pre-2013 productivity percentile & 1.117 & 1.332 & $0.445^{***}$ & $-0.809^{***}$ (0.089) & $<.001$ \\
Pre-2013 impact percentile & 0.480 & 0.528 & $0.844^{**}$ & $-0.169^{**}$ (0.052) & .001 \\
Pre-2013 coauthor status & 0.248 & 0.267 & 0.964 & $-0.037$ (0.052) & .484 \\
Academic age in 2012 & 13.257 & 15.126 & $0.888^{*}$ & $-0.118^{*}$ (0.053) & .026 \\
\bottomrule
\end{tabular}
\par\smallskip
\begin{minipage}{0.9\textwidth}\footnotesize
\textit{Notes.} Columns 2 and 3 are unconditional comparisons of raw means; columns 4 to 6 report a single multivariate exit logit with all five predictors, z-scored, entered together. Standard errors in parentheses. $^{*}p<.05$, $^{**}p<.01$, $^{***}p<.001$.
\end{minipage}
\end{table}
}

\likai{\textbf{Why the direction of selection matters.} For survivorship to manufacture the rigidity penalty, selection would have to retain scientists who persisted \textit{and} succeeded while removing scientists who adapted \textit{and} failed. The observed selection runs the other way: the scientists who disappear are those with weak prior records, regardless of how persistent they were. Venue-external citation data support this reading. Figure~\ref{fig:openalex_leavers} counts citations to the cohort's pre-2013 papers from anywhere in OpenAlex, not only from these conferences. Leavers' papers receive fewer citations than stayers' papers in every post-2013 year. This also speaks to the concern that our outcome is defined entirely within the observed venue system.}
%, and the two normalized trajectories run broadly parallel rather than diverging
\likai{
\begin{figure}[htbp]
\centering
\includegraphics[width=0.75\textwidth]{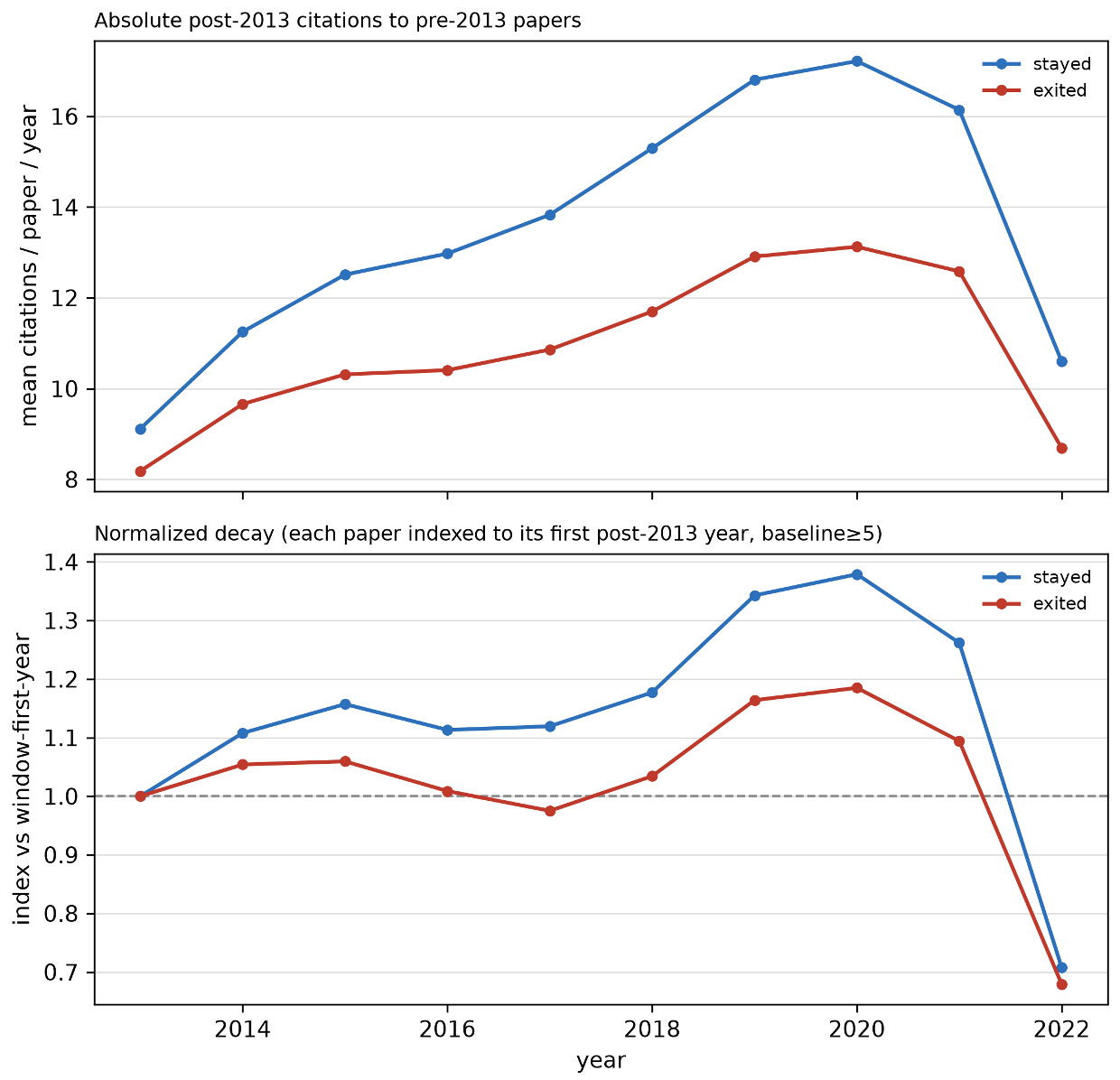}
\footnotesize
\caption{OpenAlex citations to the cohort's pre-2013 papers, leavers versus stayers. Upper panel: mean citations per paper per year, counted from anywhere in OpenAlex rather than only within the conferences. Lower panel: the same series with each paper indexed to its first post-2013 year, restricted to papers with a baseline of at least five citations. The 2022 decline likely reflects incomplete citation accrual in the final year for both groups.}
\label{fig:openalex_leavers}
\end{figure}
}

\likai{\textbf{Bounds.} We bound the estimate by returning exiters to the panel under different imputation rules and re-estimating the main impact model (Table~\ref{tab:bounds}). Under a neutral rule in which each exiter continues at their own pre-period level, the penalty survives and is slightly stronger ($-0.177$ to $-0.193$, $p<.001$). Under two worst-case rules, in which every exiter is recorded at zero impact or at the within-group first percentile, the coefficient is small and indistinguishable from zero. We report these rows rather than omit them. They impose an assumption, namely that departing scientists had literally no impact anywhere, which is inconsistent both with the exit model in Table~\ref{tab:exit} and with the venue-external citation evidence in Figure~\ref{fig:openalex_leavers}; we therefore regard them as extreme bounds rather than plausible scenarios.}

\likai{
\begin{table}[htbp]
\centering
\footnotesize
\caption{Survivorship bounds on the post-2013 $\times$ persistence coefficient (impact).}
\label{tab:bounds}
\begin{tabular}{lccc}
\toprule
Specification & Post $\times$ persistence & $p$ & $N$ \\
\midrule
Baseline (survivors only) & $-0.1771^{***}$ (0.0516) & $<.001$ & 11,016 \\
Neutral imputation: exiters carry own pre-period level & $-0.1929^{***}$ (0.0401) & $<.001$ & 17,420 \\
Worst case: exiters at zero impact & 0.0547 (0.0484) & .258 & 17,420 \\
Worst case: exiters at within-group 1st percentile & 0.0350 (0.0478) & .464 & 17,420 \\
%Neutral imputation, matched subsample & $-0.1840^{***}$ (0.0428) & $<.001$ & 12,577 \\
%Worst case at zero, matched subsample & 0.0254 (0.0495) & .608 & 12,577 \\
%Worst case at 1st percentile, matched subsample & 0.0183 (0.0492) & .710 & 12,577 \\
\bottomrule
\end{tabular}
\par\smallskip
\begin{minipage}{0.9\textwidth}\footnotesize
\textit{Notes.} Exiters are returned to the panel under different imputation rules and the model is re-estimated. The neutral rule assigns each exiter their own pre-period level; the two worst-case rules assign zero impact or the within-group first percentile. Standard errors in parentheses. $^{*}p<.05$, $^{**}p<.01$, $^{***}p<.001$.
\end{minipage}
\end{table}
}

\likai{\textbf{Tracing exiters through preprints.} A natural complement would be to follow exiters' output in preprints that never appeared at a conference. We ran a feasibility probe on 40 randomly sampled exiters. arXiv's author search resolves only surname plus initial and returns no ORCID or affiliation: 67.5\% of queries returned three or more distinct researchers, and the surname agreement rate was only 0.16. Mapping preprint output to specific authors at scale would therefore produce a match quality far below that of our disambiguated bibliographic data (Appendix~\ref{subsec:name_disambiguation}), and we do not pursue it.} %We would revisit this analysis if arXiv metadata with reliable author identifiers become available.}

\newpage
  
\section{\texorpdfstring{Topic popularity and the rigidity penalty}{Topic popularity and the rigidity penalty}}
\label{topic_nomal}

\likai{A central threat to interpretation is that persistence may co-move with topic decline. A scientist who continues working on, for example, kernel methods will register high persistence while their citations fall as the field moves toward deep learning, without persistence itself playing any role. Because persistence is defined relative to the scientist's own prior work (years before $T$) rather than to other research in the field, it does not mechanically decline as the field moves on; the concern is that it may co-move with topic decline. A related concern is that our outcome is defined within ICML and NeurIPS, so the penalty might reflect these venues' growing preference for deep learning topics rather than anything about the scientist. We address both concerns by measuring the topic channel directly and testing whether the persistence coefficient survives net of it.}

\likai{\textbf{Topic model.} We fit a 30-topic LDA model on the ICLR-inclusive corpus. Each paper is assigned the topic distribution estimated by the model, and each author-year is assigned its dominant topic (the topic with the highest weight). Across all analyses the baseline coefficient ($-0.154$) is reported in the first row of each table.}

\likai{\textbf{Topic fixed effects and topic-normalized citations.} Table~\ref{tab:topic_robust} reports two specifications. Adding author-year dominant-topic fixed effects moves the impact coefficient to $-0.167$ ($p<.001$), in the opposite direction from what the confound would predict. Replacing the outcome with a topic-normalized citation percentile, computed among papers on the same topic in the same year, yields $-0.124$ ($p=.011$), roughly a 20\% attenuation. If the penalty merely reflected which topics the venues favored, within-topic normalization should eliminate it. It does not: persistent scientists are cited less than other scientists working on the same topic, in the same year, at the same venue. The productivity coefficient is unchanged by topic fixed effects.}

\likai{
\begin{table}[htbp]
\centering
\footnotesize
\caption{Topic robustness: topic fixed effects and topic-normalized citations.}
\label{tab:topic_robust}
\begin{tabular}{lcccc}
\toprule
Specification & Post $\times$ persistence & $p$ & $N$ & $R^2$ \\
\midrule
Impact: baseline & $-0.154^{**}$ (0.051) & .002 & 11,649 & 0.0297 \\
Impact: + topic fixed effects & $-0.167^{***}$ (0.050) & $<.001$ & 11,649 & 0.0397 \\
Impact: topic-normalized citation percentile & $-0.124^{*}$ (0.049) & .011 & 11,657 & 0.0265 \\
\quad (same topic, same year) & & & & \\
Productivity: baseline & $0.0079^{***}$ (0.0023) & $<.001$ & 12,025 & 0.7486 \\
Productivity: + topic fixed effects & $0.0079^{***}$ (0.0023) & $<.001$ & 12,025 & 0.7510 \\
\bottomrule
\end{tabular}
\par\smallskip
\begin{minipage}{0.9\textwidth}\footnotesize
\textit{Notes.} Two-way fixed effects, author-clustered standard errors, and controls as in the main specification. Topic fixed effects are author-year dominant-topic fixed effects from the 30-topic LDA model. Standard errors in parentheses. $^{*}p<.05$, $^{**}p<.01$, $^{***}p<.001$.
\end{minipage}
\end{table}
}

\likai{\textbf{Topic momentum.} We next price the mechanical channel directly. For each author-year, \textit{topic momentum} is the year-over-year change in the conference share of the topics that the author actually worked on that year. Adding it to the main regression (Table~\ref{tab:momentum}) shows that topic momentum is itself a strong predictor of impact ($+0.67$, $p=.001$): the channel described above is real. But the persistence coefficient for impact barely moves, from $-0.154$ to $-0.152$ ($p=.003$), and the persistence coefficient and its significance for productivity do not change.}

\likai{\textbf{Timing structure.} We also estimate a specification in which persistence at $T$ predicts impact at $T+2$, with topic momentum at $T+1$ controlled. The coefficient is $-0.043$ ($p=.579$). We report this as an uninformative estimate rather than as evidence of no association. Requiring an author to be observed at $T$, $T+1$, and $T+2$ is a severe constraint in a panel where scientists do not publish every year: the sample falls from 11,649 to 3,773 author-years, the standard error rises from 0.051 to 0.077, and the restriction drops precisely the authors whose adaptation is most visible.}

\likai{
\begin{table}[htbp]
\centering
\footnotesize
\caption{Topic momentum and the $T+2$ timing specification.}
\label{tab:momentum}
\begin{tabular}{lcccc}
\toprule
Specification & Post $\times$ persistence & $p$ & $N$ & $R^2$ \\
\midrule
Impact: baseline & $-0.154^{**}$ (0.051) & .002 & 11,649 & 0.0297 \\
Impact: + topic momentum & $-0.152^{**}$ (0.051) & .003 & 11,649 & 0.0315 \\
Impact at $T+2$ on persistence at $T$, & $-0.043$ (0.077) & .579 & 3,773 & 0.0124 \\
\quad momentum at $T+1$ controlled & & & & \\
Productivity: baseline & $0.0079^{***}$ (0.0023) & $<.001$ & 12,025 & 0.7486 \\
Productivity: + topic momentum & $0.0079^{***}$ (0.0023) & $<.001$ & 12,025 & 0.7486 \\
\bottomrule
\end{tabular}
\par\smallskip
\begin{minipage}{0.9\textwidth}\footnotesize
\textit{Notes.} Topic momentum is the year-over-year change in the conference share of the topics the author worked on that year; its own coefficient on impact is $+0.67$ ($p=.001$). The $T+2$ row retains roughly one third of the data and is imprecise. Standard errors in parentheses. $^{*}p<.05$, $^{**}p<.01$, $^{***}p<.001$.
\end{minipage}
\end{table}
}

\likai{Taken together, topic popularity accounts for roughly a fifth of the rigidity penalty, not the whole of it. What remains is attributable to a scientist's position relative to their own prior work rather than to the popularity of the topic they work on. The component of the penalty that reflects reviewers' preferences, as distinct from genuine differences in quality, cannot be separated with these data; we discuss this in Section~\ref{section7}.}

\newpage
\section{\texorpdfstring{Complementary causal analysis}{Complementary causal analysis}}
\label{causal}

\likai{This appendix reports four analyses that complement the two-way fixed-effects panel model: an event study, a selection-on-observables matching analysis, a quasi-experimental heterogeneity analysis of pre-shock proximity to deep learning, and formal tests of \revtwo{the shape of returns to persistence}. None of these delivers an identified causal effect; we label each by the assumptions it rests on.}

\subsection{\texorpdfstring{Event study}{Event study}}
\label{app:event}

\likai{We replace the single post-2013 $\times$ persistence term with year-specific interactions, using 2012 as the base year and leaving the rest of the model unchanged (Table~\ref{tab:event}). Pre-2013 coefficients are jointly insignificant (impact: $p=.347$; productivity: $p=.505$), so we find no evidence that high- and low-persistence scientists were already diverging before the shift. The impact penalty is near zero in 2013--2015, turns negative thereafter, and reaches $-0.276$ ($p=.014$) by 2022, indicating that the penalty accumulated over time rather than appearing at once. The difference between the mean post- and pre-period coefficients ($-0.184$) is consistent with the pooled estimate in Table~\ref{impact_table} ($-0.177$). The positive main association between persistence and productivity is unaffected.}

\likai{
\begin{table}[htbp]
\centering
\footnotesize
\caption{Event study: year-specific association of persistence with outcomes, base year 2012.}
\label{tab:event}
\begin{tabular}{ccc@{\hspace{2.5em}}ccc}
\toprule
Year & Impact & Productivity & Year & Impact & Productivity \\
\midrule
2004 & 0.265 (0.203) & $-0.0034$ (0.0059) & 2013 & $-0.034$ (0.115) & 0.0000 (0.0038) \\
2005 & 0.027 (0.178) & $-0.0068$ (0.0052) & 2014 & 0.048 (0.120) & 0.0009 (0.0054) \\
2006 & 0.195 (0.176) & $-0.0085$ (0.0052) & 2015 & $-0.009$ (0.122) & 0.0009 (0.0041) \\
2007 & $-0.211$ (0.161) & $-0.0067$ (0.0043) & 2016 & $-0.163$ (0.118) & $-0.0018$ (0.0041) \\
2008 & $-0.004$ (0.147) & $-0.0039$ (0.0039) & 2017 & $-0.041$ (0.117) & $-0.0029$ (0.0040) \\
2009 & 0.146 (0.136) & $-0.0057$ (0.0046) & 2018 & $-0.120$ (0.112) & 0.0013 (0.0041) \\
2010 & 0.106 (0.136) & $-0.0026$ (0.0040) & 2019 & $-0.107$ (0.109) & 0.0014 (0.0039) \\
2011 & 0.157 (0.131) & 0.0011 (0.0051) & 2020 & $-0.169$ (0.109) & 0.0088 (0.0045) \\
2012 & \multicolumn{2}{c}{base year} & 2021 & $-0.211$ (0.110) & 0.0085 (0.0050) \\
 & & & 2022 & $-0.276$ (0.112) & 0.0025 (0.0049) \\
\bottomrule
\end{tabular}
\par\smallskip
\begin{minipage}{0.9\textwidth}\footnotesize
\textit{Notes.} Coefficients on year $\times$ persistence interactions; all other terms as in the main specification. Joint test of pre-2013 coefficients: impact $p=.347$, productivity $p=.505$.
\end{minipage}
\end{table}
}

\subsection{\texorpdfstring{Quasi-experimental heterogeneity analysis}{Quasi-experimental heterogeneity analysis}}
\label{app:threshold}

\likai{We treat the sudden rise of deep learning as a shock and compare scientists whose pre-shock work was closer to or farther from the coming paradigm.}

\likai{\textit{Measuring proximity.} We first identify typical deep learning work. For every paper abstract in the corpus, we compute the share of its words that are deep learning terms. We use a keyword dictionary comprising "deep network," "neural network," "convolution," "LSTM," "transformer," "attention," "CNN," "RNN," "backpropagation," "dropout," "ReLU," "encoder," "GAN," "ResNet," and "BERT," along with their full forms, to identify deep learning–related publications. We take the top 10\% of papers by this share as the clearest examples of deep learning research. We then normalize the SPECTER2 embeddings of these papers and average them across 768 dimensions to obtain a deep-learning centroid. A centroid represents where this body of work sits in semantic space, rather than where a single keyword or landmark paper happens to fall. Each author's pre-shock proximity is the cosine similarity between the centroid of their 2003--2012 papers and the deep-learning centroid.}

\likai{\textit{Local comparison.} We compare authors locally around the median of this proximity index: those between the 25th and 50th percentiles (just below the median) with those between the 50th and 75th percentiles (just above it). Scientists whose pre-shock work was already closer to deep learning show significantly higher post-2013 productivity (local difference $+0.157$, $p=.019$) and marginally higher persistence ($+0.012$, $p=.062$): they were already working on the winning topics and had less need to adapt. The difference in impact is not significant (Table~\ref{tab:threshold}).}

\likai{\textit{What this design is not.} The running variable is smoothly distributed with no natural discontinuity, so a sharp regression-discontinuity design is not available. We therefore present this as a quasi-experimental heterogeneity analysis and do not claim a regression-discontinuity estimate.}

\likai{
\begin{table}[htbp]
\centering
\footnotesize
\caption{Local comparison around the median of pre-shock proximity to deep learning.}
\label{tab:threshold}
\begin{tabular}{lcccc}
\toprule
Outcome & Local mean below median & Local mean above median & Local difference & $p$ \\
\midrule
Post-2013 impact (0--1 scale) & 0.480 & 0.492 & $+0.012$ & .356 \\
Post-2013 productivity & 1.496 & 1.654 & $+0.157^{*}$ & .019 \\
Post-2013 persistence (0--1 scale) & 0.367 & 0.379 & $+0.012$ & .062 \\
\bottomrule
\end{tabular}
\par\smallskip
\begin{minipage}{0.9\textwidth}\footnotesize
\textit{Notes.} Pre-shock proximity is the cosine similarity between the SPECTER2 centroid of the author's 2003--2012 papers and the deep-learning centroid. Below median: 25th--50th percentiles; above median: 50th--75th percentiles. $^{*}p<.05$.
\end{minipage}
\end{table}
}

\subsection{\texorpdfstring{Propensity-score matching}{Propensity-score matching}}
\label{app:psm}

\likai{We compare scientists who looked similar before 2013 but responded differently after deep learning arrived. If the two groups started out alike on observable characteristics, later differences between them are less likely to reflect where they already stood before the shift, although unobserved differences cannot be ruled out.}

\likai{\textit{Sample and groups.} The sample consists of post-2013 survivors: cohort scientists (at least one ICML or NeurIPS paper in 2003--2012) who continued to publish at these venues after 2013. For each scientist, we average their yearly persistence scores after 2013. Scientists above the median form the high-persistence group; those below it form the low-persistence group.}

\likai{\textit{Matching.} We describe each scientist by seven pre-2013 characteristics, defined as in the main text: productivity, impact, coauthor status, number of collaborators, collaboration age, academic age, and proximity to deep learning (defined in Appendix~\ref{app:threshold}). From these we estimate a propensity score for membership in the high-persistence group and pair each high-persistence scientist with the low-persistence scientist whose score is closest, one to one, using greedy nearest-neighbor matching. This yields 926 pairs. We assess balance with the standardized mean difference (SMD), the difference in group means divided by the pooled standard deviation. Before matching, the high-persistence group was clearly different, for example in productivity (SMD $=0.45$) and collaboration age (0.31). After matching, every SMD is 0.12 or smaller, and six of seven are below 0.06 (Table~\ref{tab:psm}).}

\likai{\textit{Results.} In the matched sample, persistent scientists published about 0.84 more papers per year after 2013 ($p<.001$), consistent with the main result and not explained by the pre-2013 differences we can measure. Their impact was about one percentile point lower, in the same direction as the main result, but the difference is not statistically significant ($p=.34$). A likely reason is limited power: this design assigns each scientist a single high-or-low label for the entire post-2013 decade, whereas the panel model uses year-to-year variation within scientists.}

\likai{
\begin{table}[htbp]
\centering
\footnotesize
\caption{Matching on post-2013 persistence (high vs.\ low).}
\label{tab:psm}
\begin{tabular}{lcc}
\toprule
 & Before matching & After matching \\
\midrule
\multicolumn{3}{l}{\textit{Balance (standardized mean difference)}} \\
Pre-2013 productivity & 0.453 & 0.123 \\
Pre-2013 impact & 0.120 & 0.001 \\
Pre-2013 coauthor status & 0.156 & 0.014 \\
Pre-2013 collaborators & 0.080 & 0.049 \\
Pre-2013 collaboration age & 0.308 & 0.057 \\
Academic age & $-0.073$ & 0.008 \\
Pre-2013 proximity to deep learning & 0.000 & $-0.011$ \\
\midrule
\multicolumn{3}{l}{\textit{Outcomes (high minus low persistence)}} \\
Post-2013 productivity (papers per year) & 0.959 ($p<.001$) & 0.835 ($p<.001$) \\
Post-2013 impact (0--1 scale) & $-0.006$ ($p=.515$) & $-0.009$ ($p=.340$) \\
\bottomrule
\end{tabular}
\par\smallskip
\begin{minipage}{0.85\textwidth}\footnotesize
\textit{Notes.} Sample: post-2013 survivors, split at the median of average post-2013 persistence. Propensity scores are estimated from the seven pre-2013 characteristics listed; 926 matched pairs, greedy one-to-one nearest-neighbor matching.
\end{minipage}
\end{table}
}

\subsection{\texorpdfstring{The shape of returns to persistence}{The shape of returns to persistence}}
\label{app:shape}

\likai{Figure~\ref{fig:adapt} in the main text is descriptive. To test whether an intermediate level of persistence is associated with the highest returns, we replace the linear persistence term with ten persistence bins (0--10, \ldots, 90--100; reference bin 30--40), keeping all controls and two-way fixed effects, and estimate five post-period outcomes (Table~\ref{tab:bins}).}

\likai{\textit{Individual-level returns.} We find no interior extremum. Impact is strictly decreasing in persistence, and productivity and impact-weighted output (annual papers multiplied by the citation percentile of that year's papers) are strictly increasing, each with zero changes in the sign of the slope. Because a monotone mean does not preclude a hump in the upper tail, we separately estimate the probability of reaching the top 10\% and top 5\% of impact. Both peak at the low end of persistence. Lind--Mehlum tests \citep{lind2010with} of an inverted U give $p=1.00$ for both tail outcomes, and the corresponding U-shape tests give $p=.90$ and $p=.57$; neither shape is supported. The shape conclusion holds on the 20--80 support, which contains 93.5\% of the sample.}

\likai{
\begin{table}[htbp]
\centering
\footnotesize
\caption{Ten-bin profile of persistence across five outcomes (post-2013).}
\label{tab:bins}
\begin{adjustbox}{max width=\textwidth}
\begin{tabular}{lccccc}
\toprule
Persistence bin & Impact & Productivity & Impact-weighted & P(top 10\% & P(top 5\% \\
 & (citation percentile) & (papers/year) & output & impact) & impact) \\
\midrule
0--10 & 9.445 (6.678) & $-0.315$ (0.127) & $-0.128$ (0.118) & 0.0495 (0.1078) & 0.0236 (0.0746) \\
10--20 & 2.866 (2.055) & $-0.187$ (0.043) & $-0.064$ (0.033) & 0.0340 (0.0261) & 0.0434 (0.0228) \\
20--30 & 1.412 (1.225) & $-0.090$ (0.027) & $-0.019$ (0.021) & 0.0318 (0.0156) & 0.0085 (0.0124) \\
30--40 & 0 (ref.) & 0 (ref.) & 0 (ref.) & 0 (ref.) & 0 (ref.) \\
40--50 & $-1.786$ (1.046) & 0.199 (0.030) & 0.087 (0.023) & $-0.0204$ (0.0128) & $-0.0253$ (0.0091) \\
50--60 & $-3.017$ (1.157) & 0.538 (0.046) & 0.218 (0.034) & $-0.0504$ (0.0151) & $-0.0462$ (0.0112) \\
60--70 & $-4.296$ (1.372) & 0.945 (0.073) & 0.404 (0.053) & $-0.0705$ (0.0173) & $-0.0541$ (0.0119) \\
70--80 & $-6.851$ (1.593) & 1.807 (0.143) & 0.720 (0.094) & $-0.0740$ (0.0210) & $-0.0568$ (0.0134) \\
80--90 & $-8.541$ (2.045) & 3.430 (0.285) & 1.433 (0.158) & $-0.0602$ (0.0273) & $-0.0488$ (0.0217) \\
90--100 & $-15.369$ (8.718) & 4.685 (1.471) & 2.439 (0.981) & $-0.1846$ (0.1414) & $-0.0376$ (0.0337) \\
\midrule
Arg-max bin & 0--10 & 90--100 & 90--100 & 0--10 & 10--20 \\
Slope sign changes & 0 & 0 & 0 & 2 & 2 \\
\bottomrule
\end{tabular}
\end{adjustbox}
\par\smallskip
\begin{minipage}{0.95\textwidth}\footnotesize
\textit{Notes.} Coefficients on persistence-bin indicators relative to the 30--40 bin; controls and two-way fixed effects the same as the main specification; author-clustered standard errors in parentheses. Impact-weighted output is annual papers multiplied by the citation percentile of that year's papers. Lind--Mehlum inverted-U tests for the two-tail outcomes: $p=1.00$ and $p=1.00$; U-shape tests: $p=.90$ and $p=.57$.
\end{minipage}
\end{table}
}

\likai{\textit{Field-level composition.} Table~\ref{tab:bands} reports how post-2012 author-years, output, and top-decile author-years are distributed across four persistence bands; no regression is estimated. The 25--50 band holds 59.7\% of author-years, 42.2\% of total output, 41.7\% of impact-weighted output, and 68.1\% of all top-decile author-years. An author-clustered bootstrap over 1,000 replications places the maximum in this band with probability 1.00 for total output, 0.998 for impact-weighted output, and 1.00 for top-decile author-years.}

\likai{
\begin{table}[htbp]
\centering
\footnotesize
\caption{Distribution of post-2012 author-years, output, and top-decile author-years across persistence bands.}
\label{tab:bands}
\begin{tabular}{lccccc}
\toprule
Persistence band & Author-years & Headcount share & Output share & Impact-weighted share & Top-decile author-years \\
\midrule
0--25 & 946 & 12.6\% & 6.7\% & 6.8\% & 140 (18.6\%) \\
25--50 & 4,492 & 59.7\% & 42.2\% & 41.7\% & 512 (68.1\%) \\
50--75 & 1,786 & 23.8\% & 35.5\% & 35.6\% & 99 (13.2\%) \\
75--100 & 296 & 3.9\% & 15.5\% & 15.9\% & 1 (0.1\%) \\
\bottomrule
\end{tabular}
\par\smallskip
\begin{minipage}{0.95\textwidth}\footnotesize
\textit{Notes.} Post-2012 author-years with a persistence value. Shares only; no regression is estimated. Percentages in the last column are shares of all top-decile author-years.
\end{minipage}
\end{table}
}

\likai{\textit{Reconciling the two.} The aggregate peak is a headcount effect. The 25--50 band contributes the most because it contains the most people as well as the field's most (impact-weighted) output, not because each of them does best: its share of output lies below its share of author-years. A steadily rising per-person curve and a hump-shaped total are compatible whenever the population thins out quickly at the high end of the range, as it does here. By analogy, most of a country's wheat comes from the region with the most farmland, not the region with the highest yield per acre; the harvest total reveals where the acreage is, not where an individual farmer should plant. \revtwo{Which end of the range counts as ``best'' for an individual also depends on the outcome considered. Per-author impact is highest at the low end of persistence, whereas productivity and impact-weighted output are highest at the high end (Table~\ref{tab:bins}); no single outcome peaks in the 25--50 band. Clustering in this band is therefore consistent with scientists trading off output against impact, as well as with frictions that make larger departures costly. We discuss these explanations in Section~\ref{section7}.}}

\newpage
\section{Analysis of the general machine learning corpus}
\label{appendix:distance}

In this study, we focused on NeurIPS and ICML as a relatively controlled environment to explore whether researchers who had consistently published in top-tier conferences could maintain their success without adapting to new trends. To compare NeurIPS/ICML with other research in the same field, we analyzed 579,973 papers from the Microsoft Academic Graph (MAG) dataset categorized under "machine learning" or "deep learning," spanning the same 20-year period that marked the formation and growth of deep learning. Using the SPECTER2 model \citep{bao2025division, singh-etal-2023-scirepeval}, we generated yearly embeddings for these papers to examine the distinctions between deep learning (approximated by ICLR (2013–2022)), ICML/NeurIPS, and general machine learning papers (2003–2022). Unlike word-level models like Word2Vec \citep{mikolov2013distributed}, which fail to capture the overall semantics of abstracts, or general-purpose sentence-level models such as Sentence-BERT \citep{reimers-gurevych-2019-sentence}, which lack domain-specific adaptation, or even scientific text-focused models like SciBERT \citep{reimers-gurevych-2019-sentence2}, which do not incorporate citation information, SPECTER2 offers a distinct advantage. Pre-trained on large-scale scientific literature, SPECTER2 uses an encoder-only transformer architecture with self-attention to capture contextual relationships, enhanced by citation-based contrastive learning to encode domain-specific semantic links such as thematic similarity. These features make SPECTER2 exceptionally well-suited for generating nuanced sentence-level embeddings in scientific contexts, outperforming most existing representation methods \citep{singh-etal-2023-scirepeval}. \likai{The same model underlies the scientist-level persistence measure in the main text.}

To prepare the data, we tokenized the full titles and abstracts of papers with a maximum length of 512 tokens—sufficient to cover the vast majority of content—and processed the data in batches of 32. The [CLS] token embedding from the final transformer layer was used as a pooled representation, effectively capturing the semantic meaning at the abstract level.

\begin{figure}[htbp]
    \centering
    \includegraphics[width=0.8\textwidth]{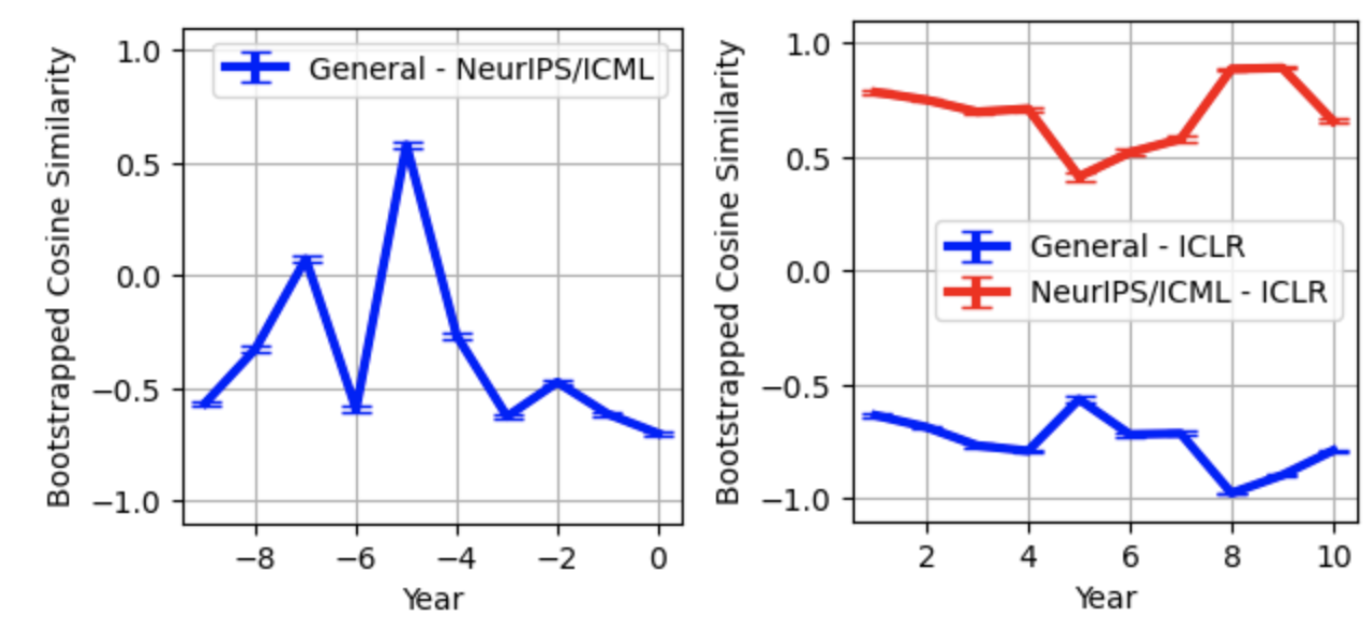} % full width
    \footnotesize
    \caption{The relationship between general machine learning research, ICLR, and ICML+NeurIPS. \likai{Distributions of pairwise SPECTER2 cosine similarity between the 500 papers nearest to each group's median embedding, by year.}}
    \label{fig:age1111}
\end{figure}

\begin{figure}[htbp]
    \centering
    \includegraphics[width=0.9\textwidth]{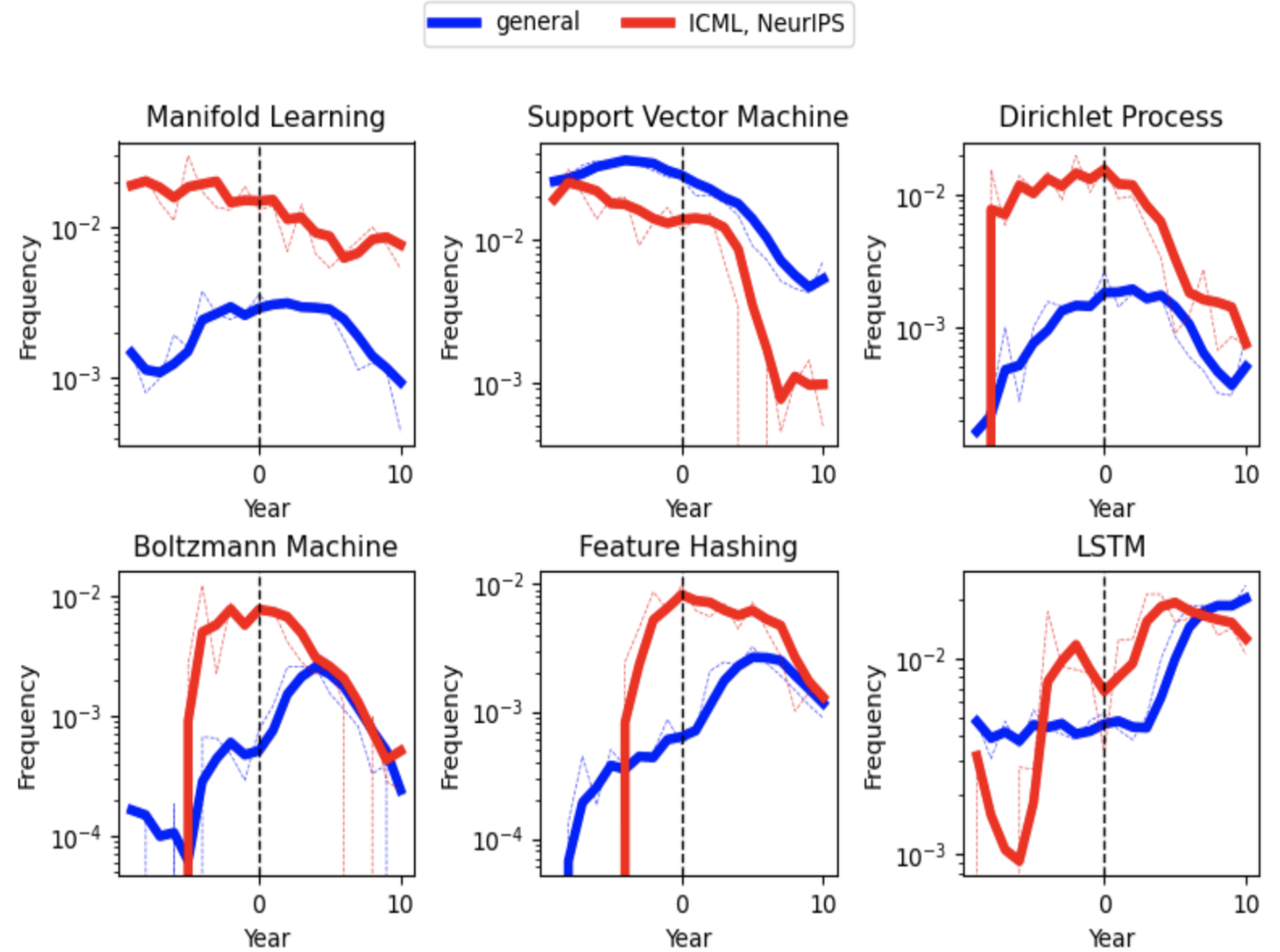} % full width
    \footnotesize
    \caption{Topical dynamics are different in ICML/NeurIPS and general machine learning. \likai{Share of abstracts mentioning each of six classic machine learning topics, by year and corpus.}}
    \label{fig:s2}
\end{figure}

We began by identifying the central embedding for ICLR, ICML/NeurIPS, and general machine learning papers, then collected the 500 nearest papers around each center to represent each cluster. Pairwise cosine similarities were calculated 5,000 times between clusters (one in 500 neighbors to one in another 500) to approximate the distribution of distances and their yearly trends. To represent the most papers within a group we use the median of each dimension of the embedding.

The results in Figure \ref{fig:age1111} reveal several notable insights. Before the rise of deep learning, NeurIPS and ICML papers already showed limited similarity to general machine learning papers (left). Following the establishment of ICLR, as anticipated, NeurIPS and ICML became increasingly aligned with ICLR, while other machine learning papers drifted further away, despite all being part of the broader machine learning field.

What makes ICML and NeurIPS special? We tracked the proportion of six classic traditional machine learning topics appearing in the abstracts of each group. As shown in Figure \ref{fig:s2}, we found that trends in ICML and NeurIPS seem to predict the trends of these topics in the general corpus. In other words, the decline of traditional machine learning topics occurred in top conferences before it was reflected in other texts. This is also why, in the main text, we consider these ICML/NeurIPS conference papers of a cohort of scientists — if the lower bar is set, the productivity of scientists almost always increases, and other venues are just not as sensitive as ICML/NeurIPS to the paradigm shift. By controlling a relatively stable and selective environment that well represents the frontier of deep learning research, we can examine \likai{the association between adaptation and} scientists' productivity and influence.

\newpage
  
\section{Team adaptation within the individual conference}
\label{appendix:within}

\begin{figure}[htbp]
    \centering
    \includegraphics[width=0.9\textwidth]{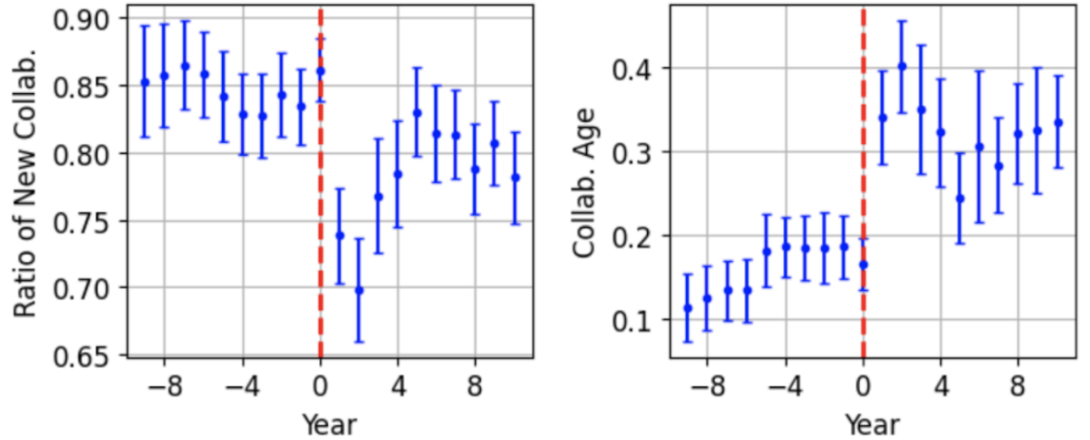} % full width
    \footnotesize
    \caption{Teams adapt through fresh collaborations within ICML. \likai{Variables defined as in Figure~\ref{fig:age2}, computed on ICML papers only.}}
    \label{fig:s3}
\end{figure}

\begin{figure}[htbp]
    \centering
    \includegraphics[width=0.9\textwidth]{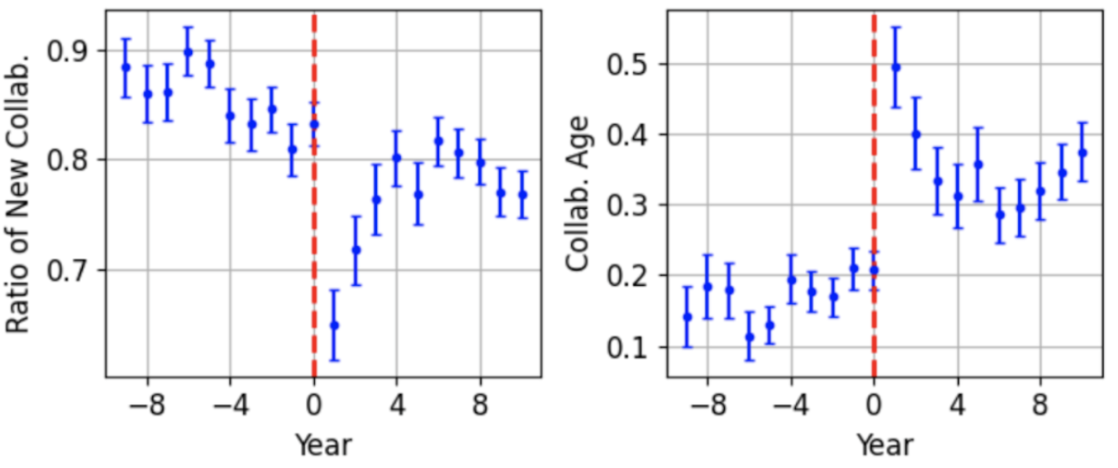} % full width
    \footnotesize
    \caption{Teams adapt through fresh collaborations within NeurIPS. \likai{Variables defined as in Figure~\ref{fig:age2}, computed on NeurIPS papers only.}}
    \label{fig:s4}
\end{figure}

As shown in Figures \ref{fig:s3} and \ref{fig:s4} the patterns reported in the main paper (Figure \ref{fig:age2}) are robust within individual conferences.

\newpage
\section{The ICLR-inclusive corpus}
\label{appendix:alternatives}

The establishment of ICLR may influence the cohort of scientists publishing in ICML and NeurIPS in two ways.

\begin{figure}[htbp]
    \centering
    \includegraphics[width=0.6\textwidth]{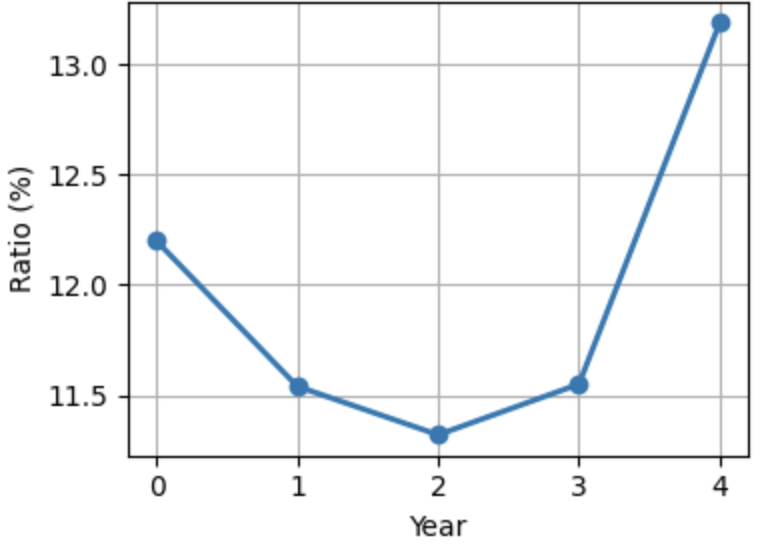} % full width
    \footnotesize
    \caption{The ratio of new authors. \likai{Share of ICML and NeurIPS authors in each year who had not appeared in any prior year.}}
    \label{fig:s5}
\end{figure}

The first possible channel is that the establishment of ICLR in 2013 (year 1) brought more attention to ICML and NeurIPS conferences, given their shared themes, leading to heightened competition and the replacement of some scientists. However, we find that the ratio of new authors (defined as those who did not appear in any prior year) in ICML and NeurIPS in the first few years after year 1 was not significantly higher than in previous and following years (see Figure \ref{fig:s5}). 

The second possible channel is that some scientists shifted their focus to publishing at ICLR so they disappeared from ICML/NeurIPS. However, only 1.21\% (65) of the focal group of scientists submitted at least one paper to ICLR 2013, yet they accounted for 65.7\% of submissions and 77.3\% of accepted papers at the conference. During the first five years of ICLR (2013–2017) - a period marked by heightened dynamism as scientists formed new collaborations to adapt (see Figures \ref{fig:age2}, \ref{fig:s3}, and \ref{fig:s4}) - papers published in ICLR accounted for only 6.8\% of this cohort's total publications. Although ICLR later grew in \likai{prominence}, it is important to recognize that such shifts in impact take time to materialize. 

\likai{Because the earliest ICLR authors were disproportionately drawn from the most adaptive members of our cohort, excluding ICLR could nonetheless bias the estimated rigidity penalty. We therefore re-estimate the main specification on an ICLR-inclusive panel in which each author-year's productivity, impact, and persistence are computed from the cohort's ICML, NeurIPS, and ICLR papers. The post-2013 $\times$ persistence coefficient is $-0.154$ for impact (SE $=0.051$, $p=.002$) and $0.0079$ for productivity (SE $=0.0023$, $p<.001$), compared with $-0.177$ and $0.005$ in the ICML/NeurIPS panel (Tables~\ref{impact_table} and \ref{prod_table}). Including ICLR thus leaves the main conclusions unchanged, although the impact penalty is modestly smaller in magnitude.}

\likai{Adaptive researchers do experience \revtwo{lower productivity}, for several possible reasons. One is a shift in attention: adaptive authors may send papers to more adaptive venues such as ICLR. Given the modest share of the cohort's output that appeared at ICLR in its early years, however, the more likely primary factor is the time required \revtwo{to learn} new methods and \revtwo{adjust} to the emerging paradigm. \revtwo{Including ICLR papers leaves the productivity coefficient positive ($0.0079$, $p<.001$).} \revtwo{This productivity cost of adaptation is also what makes the choice of how far to move a genuine trade-off between output and impact (Section~\ref{5.2.3}).} What we find most notable is that even within a publication environment these scientists were already familiar with, one that actively tries to capture the cutting edge, persistence was associated with a citation penalty.}

\newpage
\section{Relationships between variables}
\label{appendix:rela}

\begin{figure}[htbp]
    \centering
    \includegraphics[width=0.7\textwidth]{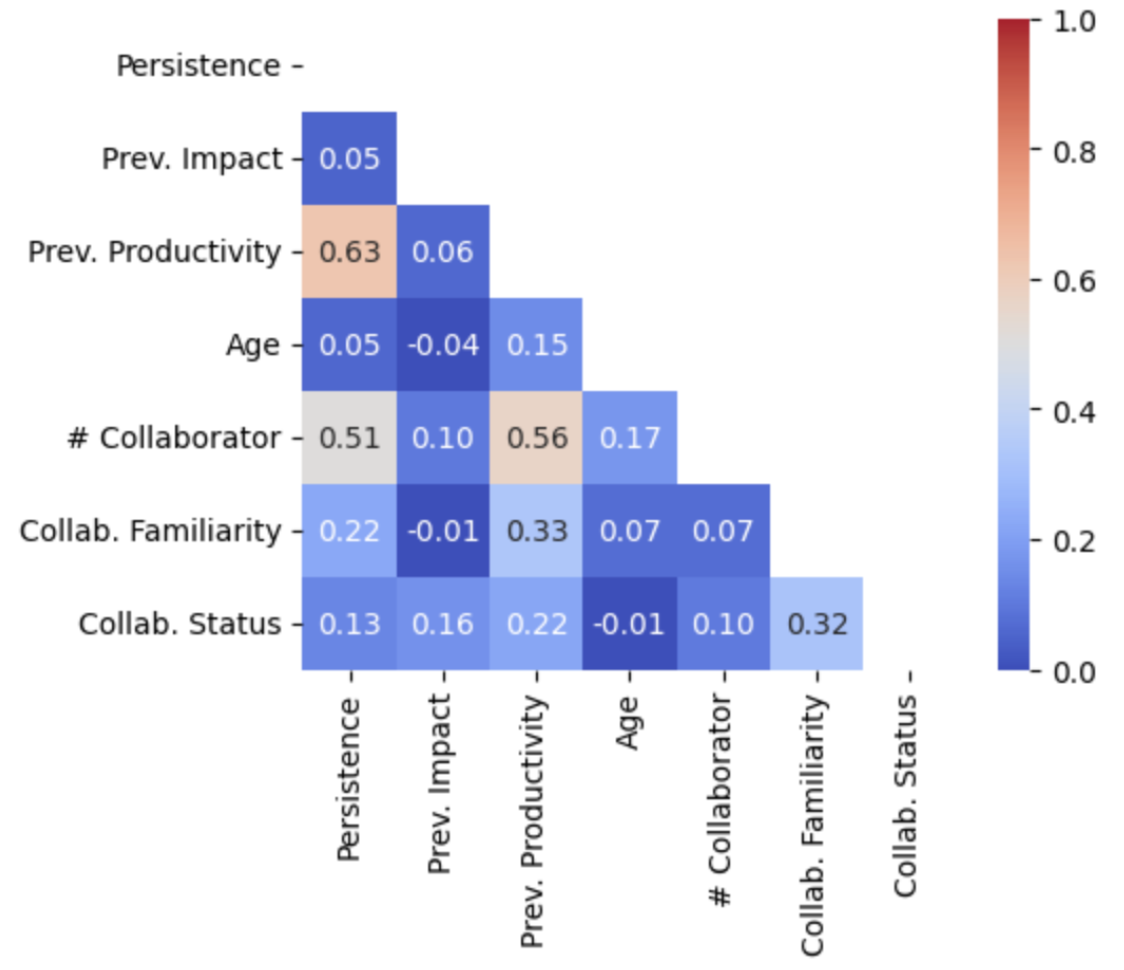} % full width
    \footnotesize
    \caption{Relationship between previous status, collaboration networks, and age. \likai{Pairwise correlations among persistence, prior impact, prior productivity, academic age, collaborator count, collaborator status, and collaboration familiarity; variables defined as in Section~\ref{6.2.1}.}}
    \label{fig:s6}
\end{figure}

To clarify the relationships between the variables, Figure \ref{fig:s6} presents the pairwise correlations among scientists' research persistence, prior impact, prior productivity, age, and features of their collaboration networks (including collaborator count, collaborator status, and collaborator familiarity). Overall, the correlations are relatively weak. There is a moderate correlation between research persistence, previous productivity, and current collaborators, but we find no evidence of multicollinearity.

\newpage 
\section{Academic age and output}
\label{appendix:age_appendix}

To what extent are the declines in productivity and impact simply natural consequences of aging? In this section, we argue that, within our study population, age is not a primary factor \likai{associated with} research output.

\begin{figure}[htbp]
    \centering
    \includegraphics[width=0.9\textwidth]{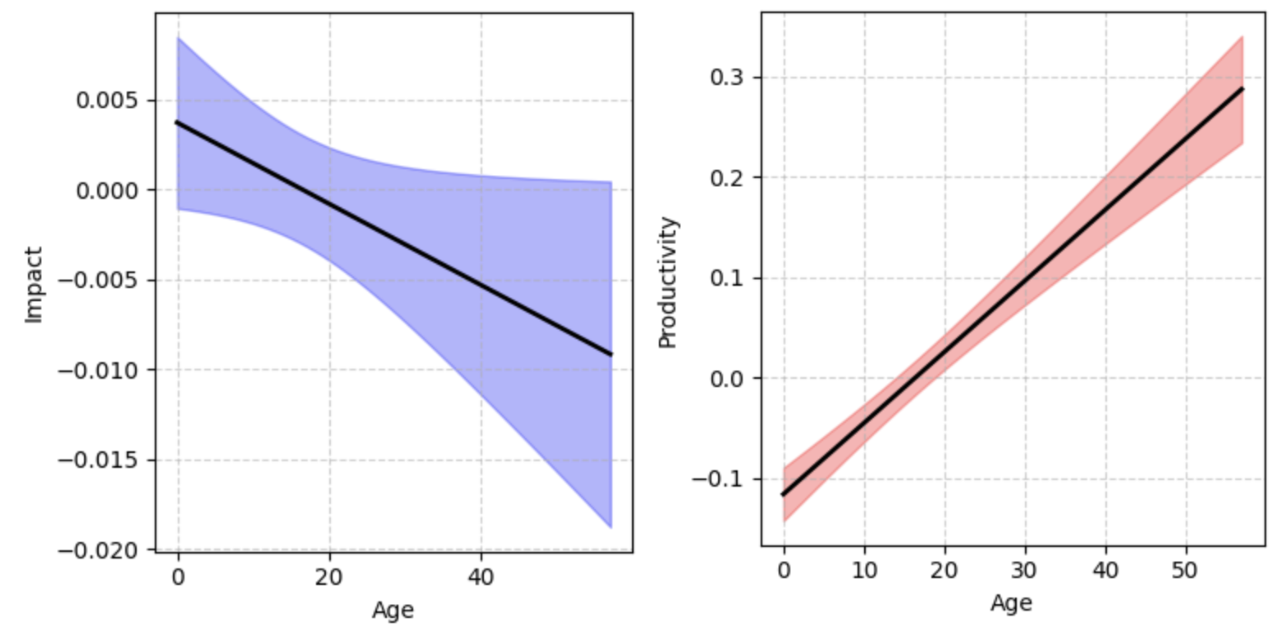} % full width
    \footnotesize
    \caption{Relationship between age and productivity/impact. \likai{Academic age is years since the scientist's first publication in OpenAlex; productivity and impact defined as in Section~\ref{6.2.1}.}}
    \label{fig:s7}
\end{figure}

Impact has a negative relation with age but the association is very weak (coefficient $=-0.0002$ (on a 0-100 scale), p = 0.048). Productivity actually has a strong positive relation with age (coefficient = 0.007, p = 0.000). Here age is at the scientist level not the team level.

\newpage 
\section{\texorpdfstring{Name disambiguation}{Name disambiguation}}
\label{subsec:name_disambiguation}

\likai{Name disambiguation is particularly crucial for computer science papers because of the high prevalence of common Chinese names \citep{kim2023effect}. The scientists in our sample published at least one paper at ICML or NeurIPS between 2003 and 2012, a period during which Chinese scientists were less represented at these conferences than they are today, which reduces but does not eliminate the problem.}

\likai{\textbf{Procedure.} We use names, email addresses, and affiliations for initial disambiguation. This approach is limited when individuals change email addresses or institutions, for example after graduating. To improve accuracy, we further use the Google Scholar API (Scholarly): if two author records with different email addresses or affiliations share the same name, and their publications on Google Scholar indicate that they are the same individual, we merge them into a single author record.}

\likai{\textbf{Validation.} Prior work benchmarking OpenAlex against large-scale gold-standard disambiguation datasets reports relatively strong performance, with an F1-score of approximately 0.82 \citep{zhang2023lagos}. Here we randomly sampled 100 author names and IDs and checked them manually, finding a 96\% match rate. A match means that the papers attributed to the author in the OpenAlex data are, with certainty, their own. We manually improved the raw performance to a much higher accuracy.}

\newpage 
\section{Examples of TF-IDF token extraction}
\label{word_table}

\begin{table}[ht]
    \centering
    \begin{adjustbox}{max width=\linewidth}
    \begin{tabularx}{\linewidth}{|>{\raggedright\arraybackslash}X|>{\raggedright\arraybackslash}X|}
        \hline
        \textbf{Raw Abstract} & \textbf{10 Extracted Tokens and Their TF-IDF Scores} \\
        \hline
        When we learn a new motor skill, we have to contend with both the variability inherent in our sensors and the task. The sensory uncertainty can be reduced by using information about the distribution of previously experienced tasks. Here we impose a distribution on a novel sensorimotor task and manipulate the variability of the sensory feedback. We show that subjects internally represent both the distribution of the task as well as their sensory uncertainty. Moreover, they combine these two sources of information in a way that is qualitatively predicted by optimal Bayesian processing. We further analyze whether the subjects can represent multimodal distributions such as mixtures of Gaussians. The results show that the CNS employs probabilistic models during sensorimotor learning even when the priors are multimodal. 
        & 
        \texttt{\{sensorimotor: 0.467, sensory: 0.346, variability: 0.231, multimodal: 0.222, distribution: 0.204, cns: 0.19, subject: 0.188, represent: 0.172, task: 0.169, internally: 0.164\}} \\
        \hline
    \end{tabularx}
    \end{adjustbox}
    \label{tab:my_label}
\end{table}

\end{document}